\documentclass[
  reprint,
  superscriptaddress,
  amsmath,amssymb,
  aps,
]{revtex4-1}

\usepackage{lineno,hyperref}   
\usepackage[pdftex]{graphicx}  
\usepackage{dcolumn} 
\usepackage{bm} 

\usepackage{color}

\begin{document}  

\title{Measurement of $\gamma$-ray production via the neutron-${\bf ^{16}O}$ 
       reaction \\ using a 77~MeV quasi-monoenergetic neutron beam}
\author{Y. Ashida}
 \email{assy.8594.1207.physics@gmail.com}
 \affiliation{Department of Physics, Kyoto University, Kyoto, Japan}
\author{H. Nagata}
 \affiliation{Department of Natural Science and Technology, 
              Okayama University, Okayama, Japan}
\author{M. Mori}
 \affiliation{Department of Physics, Kyoto University, Kyoto, Japan}
\author{G. Collazuol}
 \affiliation{Department of Physics and Astronomy, 
              University of Padova, Padova, Italy}
\author{D. Fukuda}
 \affiliation{Department of Natural Science and Technology, 
              Okayama University, Okayama, Japan}
\author{T. Horai}
 \affiliation{Department of Natural Science and Technology, 
              Okayama University, Okayama, Japan}
\author{\\ F. Iacob}
 \affiliation{Department of Physics and Astronomy, 
              University of Padova, Padova, Italy}
\author{A. Konaka}
 \affiliation{TRIUMF, Vancouver, Canada}
 \affiliation{Research Center for Nuclear Physics (RCNP), Osaka, Japan}
\author{Y. Koshio}
 \affiliation{Department of Natural Science and Technology, 
              Okayama University, Okayama, Japan}
\author{T. Nakaya}
 \affiliation{Department of Physics, Kyoto University, Kyoto, Japan}
\author{C. Nantais}
 \affiliation{Department of Physics, University of Toronto, Toronto, Canada}
\author{T. Shima}
 \affiliation{Research Center for Nuclear Physics (RCNP), Osaka, Japan}
\author{\\ A. Suzuki}
 \affiliation{Department of Physics, Kobe University, Kobe, Japan}
\author{Y. Takeuchi}
 \affiliation{Department of Physics, Kobe University, Kobe, Japan}
\author{H. Tanaka}
 \affiliation{SLAC National Accelerator Laboratory, Menlo Park, California, USA}
\author{R. Wendell}
 \affiliation{Department of Physics, Kyoto University, Kyoto, Japan}
\author{T. Yano}
 \affiliation{Department of Physics, Kobe University, Kobe, Japan}
%
%
%
%
\date{\today}

\begin{abstract}

Understanding of $\gamma$-ray production via neutron interactions on oxygen 
is essential for the study of neutrino neutral-current quasielastic interactions 
in water Cherenkov detectors. 
A measurement of $\gamma$-ray production from such reactions was performed 
using a 77~MeV quasi-monoenergetic neutron beam.
Several $\gamma$-ray peaks, which are expected to come from neutron-${\rm ^{16}O}$ 
reactions, are observed and production cross sections are measured for 
nine $\gamma$-ray components of energies between 2 and 8~MeV. 
These are the first measurements at this neutron energy 
using a nearly monoenergetic beam. 

\end{abstract}

\maketitle

\section{Introduction}
\label{sec:introduction}

Precise knowledge of the neutrino neutral-current quasielastic (NCQE) interaction 
on oxygen is crucial for a variety of physics studies at water Cherenkov detectors, 
such as Super-Kamiokande (SK) \cite{superk}, the gadolinium-loaded SK (SK-Gd) \cite{skgd}, 
and Hyper-Kamiokande \cite{hyperk}. 
Indeed, the NCQE scattering of atmospheric neutrinos is one of the main background sources 
in searches for supernova relic neutrinos (SRNs) in these experiments~\cite{sksrn12,sksrn123, sksrn4, sksrn1234} 
and is similarly a background to searches for dark matter in long-baseline accelerator 
neutrino experiments~\cite{t2kdm1,t2kdm2}. 
A sample enriched in NCQE interactions can also be used to investigate the possibility of sterile neutrinos 
since its cross section does not depend on the active neutrino flavor~\cite{t2ksterile}.

Measurements of the neutrino NCQE scattering cross section in water Cherenkov detectors 
can be made by observing de-excitation $\gamma$-rays emitted from nuclei recoiling from 
the interaction with a neutrino~\cite{ankowski}. 
However, this method suffers from large backgrounds in the low energy region ($E < 20$~MeV)
due to radioactive and cosmogenic events.
The T2K experiment~\cite{t2kexp} overcame this difficulty by using timing information 
from its pulsed neutrino beam to measure the NCQE interaction cross section~\cite{t2kncqe1to3,t2kncqe1to9}.
Not only is this measurement directly applicable to estimating the background 
to dark matter searches and enabling sterile neutrino searches~\cite{t2ksterile}, 
since the peak energy ($\approx$600~MeV) is near the peak of the atmospheric neutrino spectrum 
it can also be used to estimate backgrounds to SRN searches. 

\begin{figure}[htbp]
  \begin{center}
   \includegraphics[clip,width=9.0cm]{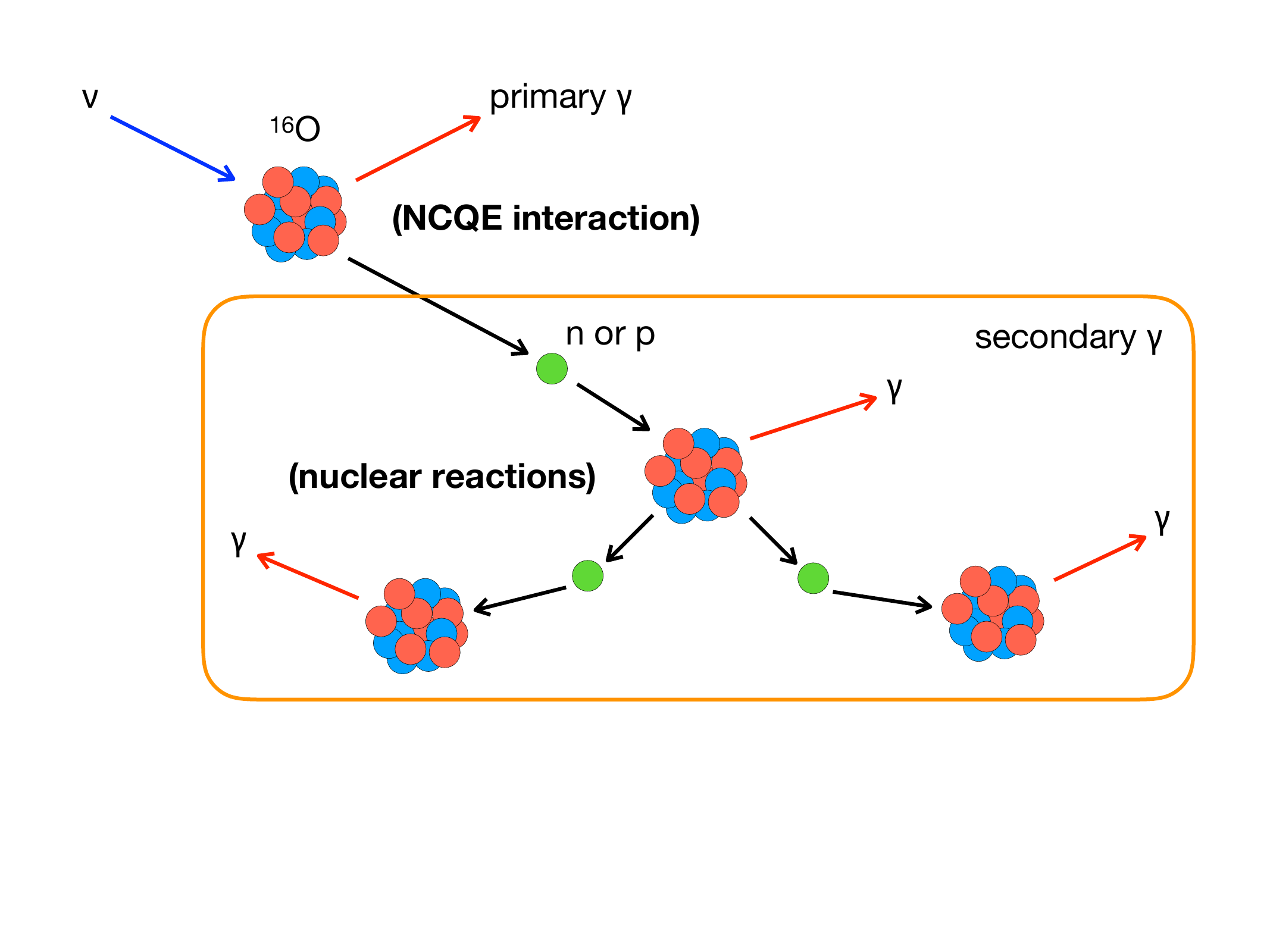}
  \end{center}
  \vspace{-55truept}
  \caption{Schematic illustration of primary and secondary $\gamma$-ray productions
           via neutrino and subsequent nuclear reactions in water.
	   The figure is taken from Ref.~\cite{t2kncqe1to9}.}
  \label{fig:ncgammaschematic}
  \end{figure}

Despite the success of this measurement, it suffers from large systematic errors due to 
the uncertainties associated with hadronic secondaries produced in the initial 
neutrino-nucleus interaction. 
Indeed, neutrino interactions at several hundreds of MeV usually knock out one or more 
nucleons with energies ranging from a few tens to several hundred MeV, which 
subsequently interact within the target material.
Protons and ions are often below the Cherenkov threshold and stop
before undergoing hadronic interactions that could produce $\gamma$-rays, and 
so their effect on the NCQE measurement is small.
Neutrons, on the other hand, interact with other nuclei inside the detector leading to 
additional $\gamma$-ray production, as shown in Figure~\ref{fig:ncgammaschematic}.  
The $\gamma$-rays from such secondary nuclear interactions are difficult to distinguish 
from those induced by the primary neutrino-nucleus interaction, since 
they have similar energies and are separated in time only by O(10)~ns.
Therefore, the T2K NCQE scattering measurement relies on Monte Carlo (MC) simulations 
to estimate the rate of secondary $\gamma$-ray production.
At present, its primary model (GCALOR~\cite{gcalor1,gcalor2}) does not reproduce 
the observed data well, which results in a large systematic uncertainty
(see Refs.~\cite{t2kncqe1to3,t2kncqe1to9} for details). 

Within GCALOR the ENDF/B-V library \cite{endf} is used to simulate neutron reactions 
below 20~MeV and an intra-nuclear cascade model is used above 20~MeV. 
While the latest version of ENDF/B, VIII, added new experimental data, the data for 
reactions above 20~MeV is limited. 
Further, the intra-nuclear cascade model is known to be insufficient for energies between 
20 and 200~MeV though it describes hadronic phenomena above 200~MeV well~\cite{inc1,inc2}. 
This difficulty is compounded by the fact that photon emission from neutron interactions above 20~MeV 
is currently based on little experimental data.
To improve the current nuclear reaction model reliable cross section measurements of these processes 
are necessary. 
The purpose of the present work is to measure $\gamma$-ray production induced by neutron-${\rm ^{16}O}$ 
reactions and thereby provide information for the development of neutron interaction models.

This paper reports the results from the E487 experiment carried out in Osaka University's 
Research Center for Nuclear Physics (RCNP) \cite{rcnp1,rcnp2,rcnp3}. 
The experimental details are given in Section~\ref{sec:experiment} and the analysis results 
are shown in Sections~\ref{sec:neutronflux} to \ref{sec:xsection}.
After discussing the measurement results in Section~\ref{sec:discussion}, 
concluding remarks are presented in Section~\ref{sec:conclusion}.

\section{Experiment}
\label{sec:experiment}

\subsection{Facility and beam properties}

\begin{figure*}[htbp]
 \begin{center} 
  \includegraphics[clip,width=12.0cm]{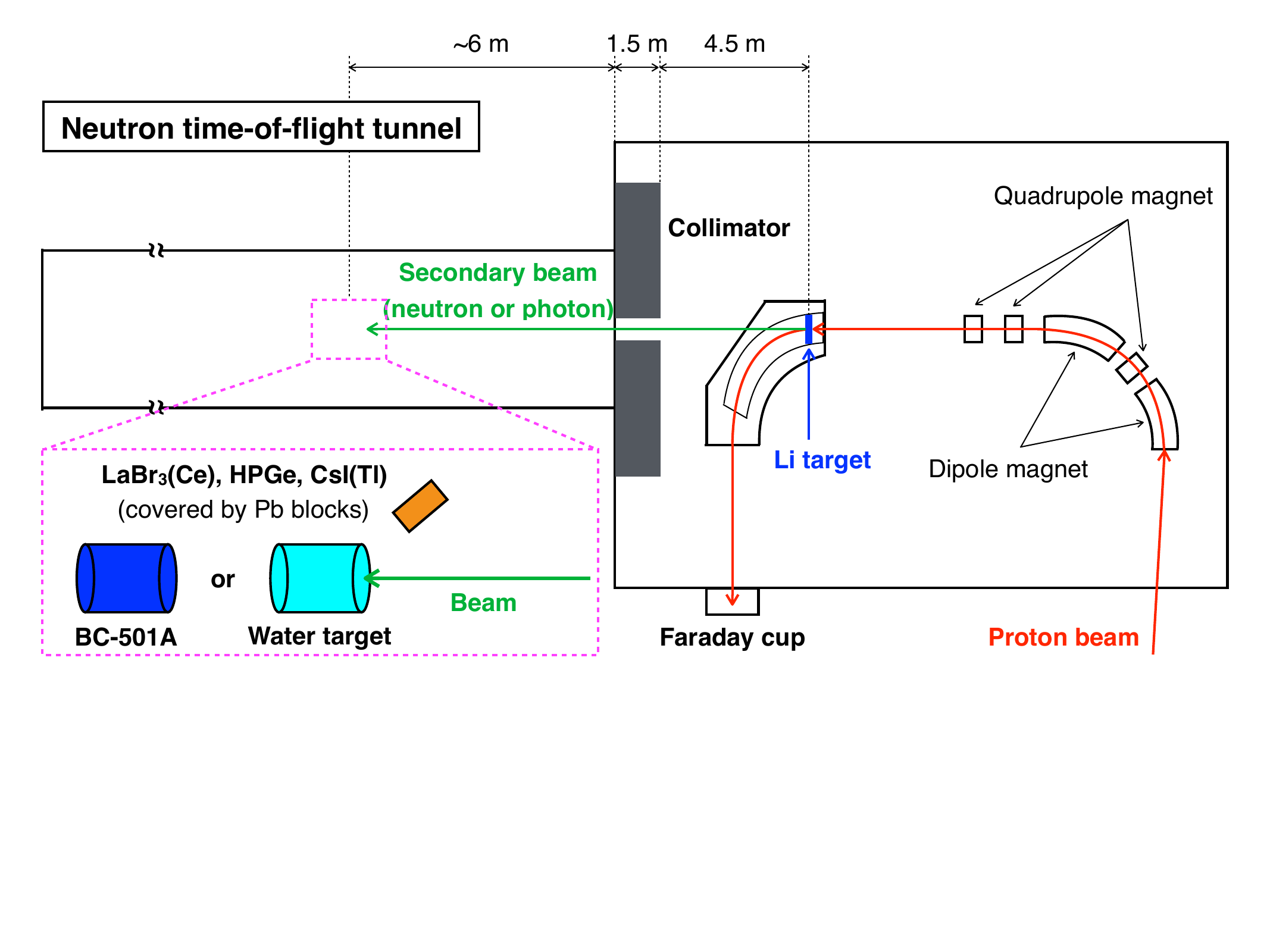}
 \end{center}
 \vspace{-80truept}
 \caption{Schematic drawing of the RCNP facility and the neutron time-of-flight beamline. 
          The dotted box shows a magnified depiction of the experimental setup.}
       
 \label{fig:rcnpfacility}
 \end{figure*}

The E487 experiment was carried out in a 100~m long neutron time-of-flight beamline at RCNP. 
A proton beam was accelerated to a kinetic energy of 80 MeV using two cyclotrons, 
the K140 AVF cyclotron and the K400 ring cyclotron, and then directed onto a 10~mm thick 
lithium target (${\rm ^{nat}Li}$: 92.5\% ${\rm ^{7}Li}$ and 7.5\% ${\rm ^{6}Li}$) to produce 
an almost monoenergetic neutron beam via the ${\rm ^{7}Li}(p,n){\rm ^{7}Be}$ reaction.
This monoenergetic beam allows for a clean measurement of the neutron interaction cross section 
at a single energy with limited contamination from neutrons of other energies. 
The proton beam size was tuned to be small compared to the lithium target size. 
During the experiment the proton energy was kept at 80 $\pm$ 0.6~MeV. 
The proton beam structure had 200~ps wide bunches separated in time by 62.5~ns 
and a chopper was used to select only one bunch in nine for neutron beam production.
After chopping the beam current was tuned from a few to 110~nA.
Downstream of the lithium target a magnetic field is used to bend charged particles 
towards a beam dump such that only neutral particles (neutrons and photons) enter the beamline.
A Faraday cup placed at the beam dump is used to measure the proton beam current. 
The 80~MeV setting is below the pion production threshold and therefore contamination of 
high energy $\gamma$-rays from neutral pion decay is expected to be negligible in the beam. 
A few particles which are not fully bent by the magnet are stopped in an iron and concrete 
collimator placed 4.5~m away from the lithium target. 
The collimator depth is 1.5~m and has an aperture of $10 \times 12$~cm$^{2}$.  
Figure~\ref{fig:rcnpfacility} shows a schematic drawing of the facility with the experimental 
setup located 12~m downstream of the lithium target. 

\subsection{Experimental setup}

A cylindrical acrylic container with a 20.0~cm diameter and a 26.5~cm length was placed 
on the beam axis and used as a sample. 
The acrylic container is 1.0~cm thick at its ends and 0.5~cm thick at its barrel walls. 
Measurements were conducted with both water and air filling its interior. 
A lanthanum bromide scintillator, Saint-Gobain B380 ${\rm LaBr_3(Ce)}$, was used 
to detect $\gamma$-rays emitted from neutron-oxygen reactions. 
The scintillator crystal is cylindrical in shape with a 4.5~cm diameter and a 4.5~cm length.
The ${\rm LaBr_3(Ce)}$ scintillator was optically coupled to a Hamamatsu H6410 photomultiplier 
tube (PMT) and its charge and time data were read out by a VME 12-bit CAEN V792N QDC (charge to digital converter) and 
a VME 12-bit CAEN V775N TDC (time to digital converter), respectively. 
It was placed upstream of the acrylic vessel to reduce backgrounds produced by scattered neutrons. 
Its arrangement was set so that its surface is 20.5~cm away from the center of the vessel,
which serves an acceptance of 0.038~sr.
To reduce backgrounds from surrounding materials the detector was 
shielded with lead bricks on all sides except for the surface viewing the acrylic container. 
A high-purity germanium detector (HPGe) was also placed upstream of the vessel to observe 
$\gamma$-rays with high precision.
This detector is an ORTEC GEM 20180-P and uses a cylindrical coaxial crystal 55~mm (46~mm) in 
diameter (length) with a hole diameter (depth) of 9.2~mm (33.4~mm). 
Spectrum data from the HPGe were read out by an MCA Kromek K102 analyzer and saved to disk using 
its proprietary software (KSpect). 
No time data were recorded for the HPGe detector.
The detector was placed in a similar position as the ${\rm LaBr_3(Ce)}$ detector 
and shielded with lead bricks. 

Apart from the main measurement with the ${\rm LaBr_3(Ce)}$ scintillator, 
dedicated measurements of the neutron beam flux and the background arising 
from neutrons scattered in the water-filled vessel were conducted. 
For the flux measurement the acrylic container was replaced with an organic liquid scintillator 
(BC-501A, Saint-Gobain 20LA32) coupled to a Hamamatsu H6527 PMT. 
The detector was set on the beam axis in order to 
measure the neutron time-of-flight (TOF) to the position of the acrylic vessel.
The scintillator is a 5~inch diameter by 8~inch long cylindrical detector and was read out 
using the same QDC and TDC modules as used for the ${\rm LaBr_3(Ce)}$. 
Backgrounds at the $\gamma$-ray detector position arising from neutrons scattered off the vessel
were measured with an OKEN CsI(Tl) crystal, whose size is $3.5\times3.5\times3.5$~${\rm cm^3}$, 
coupled to the H6410 PMT. 
A 14-bit 250~MHz CAEN DT5725 Flash-ADC (analog to digital converter) was used to record CsI(Tl) waveform data. 
Scattered neutron measurements were done in parallel with the main measurement 
for both water-filled and empty container configurations.

In all measurements, the proton beam current was monitored with the Faraday cup.
The cup was read out by an ORTEC 439 counter for the normalization described in the analysis below.
The digital acquisition system (DAQ) dead time was measured using clock pulses during the beam test with a precision better 
than 1\% and is corrected for in the cross section measurement.
It was $\approx$7\% in the $\gamma$-ray measurement by  ${\rm LaBr_3(Ce)}$ and 10$-$40\% in the flux 
measurement by BC-501A depending on the beam intensity. 
The impact of the time variance of the DAQ dead time due to the pulsed beam structure 
on the final results is at most a few percent.
Note that the contribution to the dead time from the intrinsic radioactive impurities in 
${\rm LaBr_3(Ce)}$ is small.

\subsection{Detector calibration}

Energy calibrations for the ${\rm LaBr_3(Ce)}$, HPGe, and CsI(Tl) detectors were conducted  
using the photo-absorption peaks of $\gamma$-rays from several isotopes with a maximum energy of $\approx$8~MeV.
Relative to other errors discussed below, calibration errors are small enough to be negligible 
in the cross section measurement.
The detector gain was monitored throughout the experiment and no significant fluctuations 
were observed.

Recoil electrons from Compton-scattered $\gamma$-rays produced by an ${\rm ^{22}Na}$ source 
were used to calibrate the BC-501A detector.
The scattered $\gamma$-rays were tagged by the ${\rm LaBr_3(Ce)}$ detector at 
different geometrical positions, which allows for selection of the recoil electron energy 
using the angles made by the two detectors and the source. 
Geometrical uncertainties in the positioning of the detectors produce the largest systematic 
errors in the calibration, but result in less than a 0.1\% systematic uncertainty in 
the neutron flux measurement as described in Section~\ref{sec:neutronflux}.

\section{Neutron flux}
\label{sec:neutronflux}

As described above, in order to measure the $\gamma$-ray production cross section 
a precise measurement of the neutron flux is essential.
First, neutron-like events in the BC-501A scintillator are selected 
using the pulse shape discrimination (PSD) method discussed below and their kinetic energy 
is inferred from their TOF.
The result is converted to a flux after correcting for the 
detector efficiency as calculated using the 
SCINFUL-QMD~\cite{scinfulqmd1,scinfulqmd2} simulation.

\subsection{PSD and TOF analysis}

Neutron-like events are selected based on their pulse shape and deposited energy.
For events depositing energy within the dynamic range of the QDC, a PSD parameter 
is defined as: 

  \begin{eqnarray}
   {\rm PSD \ parameter} = \frac{Q_{\rm tail} - Q_{\rm ped}}
                                {Q_{\rm total} - Q_{\rm ped}}. 
  \label{eq:psdparam}
  \end{eqnarray}
  \vspace{1truept}

\begin{figure}[htbp]
  \begin{center}
   \includegraphics[clip,width=9.0cm]{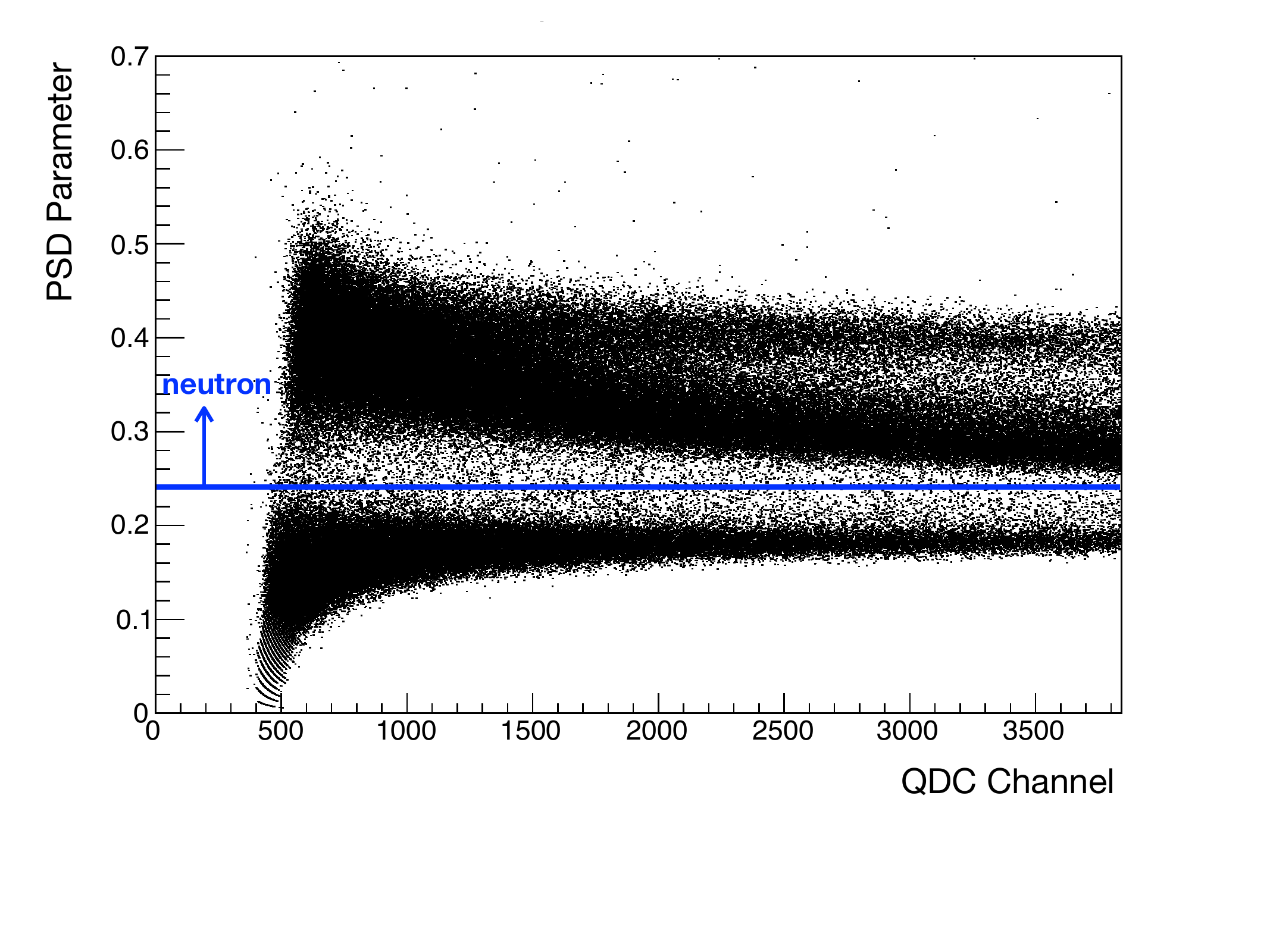}
  \end{center}
  \vspace{-40truept}
  \caption{PSD parameter as a function of the total deposited charge ($Q_{\rm total}$) value.
           The blue line represents the neutron selection criterion.}
  \label{fig:psddist}
  \end{figure}

\begin{figure}[htbp]
  \begin{center}
   \includegraphics[clip,width=9.0cm]{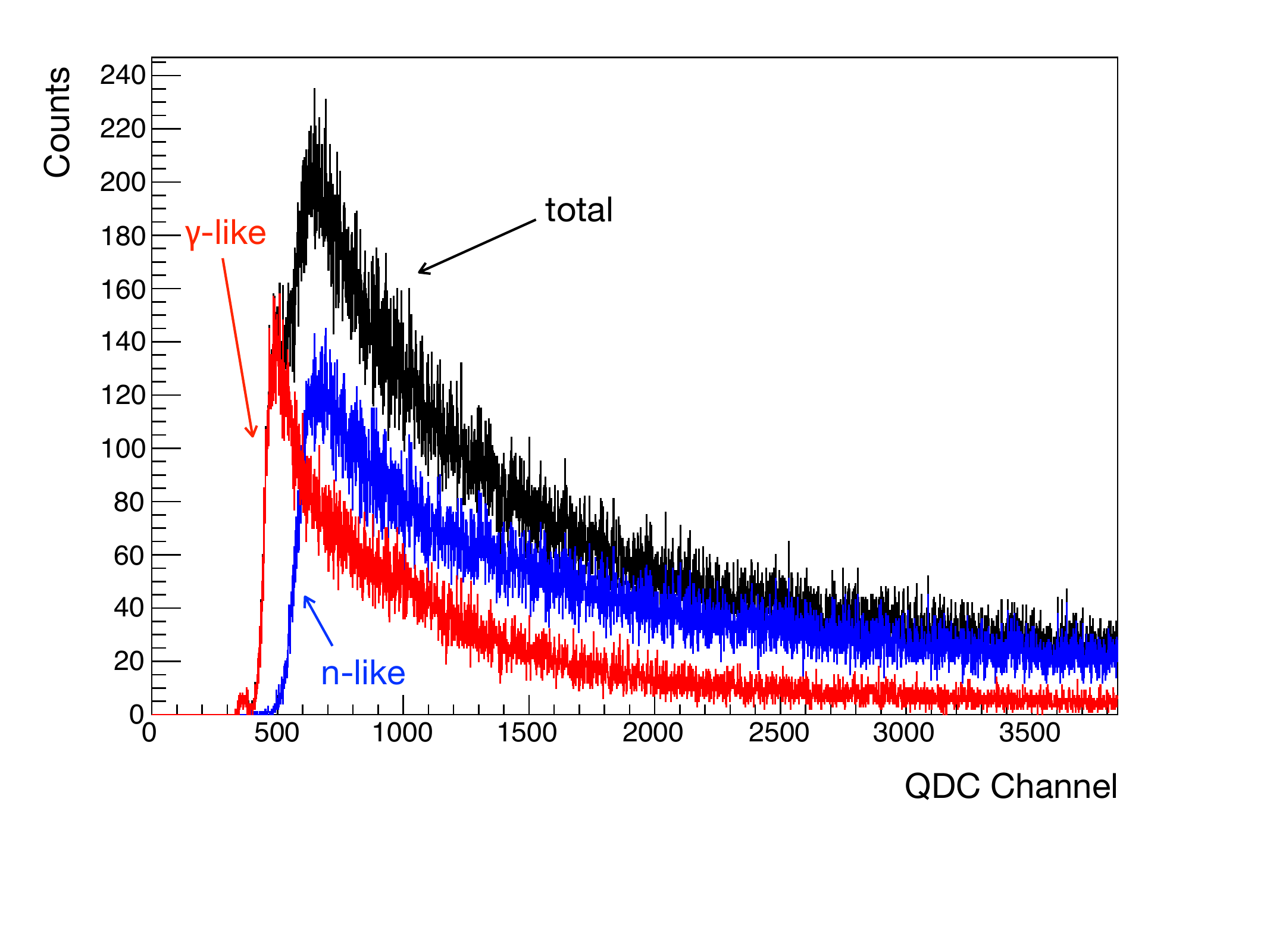}
  \end{center}
  \vspace{-40truept}
  \caption{Deposited energy in the BC-501A detector for all (black), 
           neutron-like (blue), and $\gamma$-like (red) events.}
  \label{fig:lqsspectrum}
  \end{figure}

\noindent 
Here $Q_{\rm tail}$ is the integrated charge in QDC counts of the PMT waveform for 
a pre-determined late-time window and $Q_{\rm total}$ is the charge of the entire waveform.
The $Q_{\rm ped}$ refers to an offset of the QDC module, which differs in general 
from channel to channel.
The optimal late-time integration window for $Q_{\rm tail}$ is determined by calibration 
data with neutrons from an ${\rm ^{241}Am/Be}$ source.
The distribution of the PSD parameter as a function of $Q_{\rm total}$ is shown 
in Figure~\ref{fig:psddist}. 
In this analysis events with a PSD parameter larger (smaller) than 0.24 are selected as 
neutrons ($\gamma$-rays). 
The neutron inefficiency of this cut has been confirmed to be negligible using 
an ${\rm ^{241}Am/Be}$ neutron source. 
Protons and heavier particles such as deuterons and alphas, which are induced by neutron 
interactions in the scintillator, are observed in the large PSD parameter region.
Figure~\ref{fig:lqsspectrum} shows distributions of deposited energy in the scintillator broken 
down by neutron-like and $\gamma$-like events after the PSD selection. 
Events whose energy is beyond the QDC dynamic range of $\approx$4000~channel ($\approx$6.5~MeV) 
are selected as neutrons because the contribution from $\gamma$-rays in this region is expected 
to be small.

\begin{figure}[htbp]
  \begin{center}
   \includegraphics[clip,width=9.0cm]{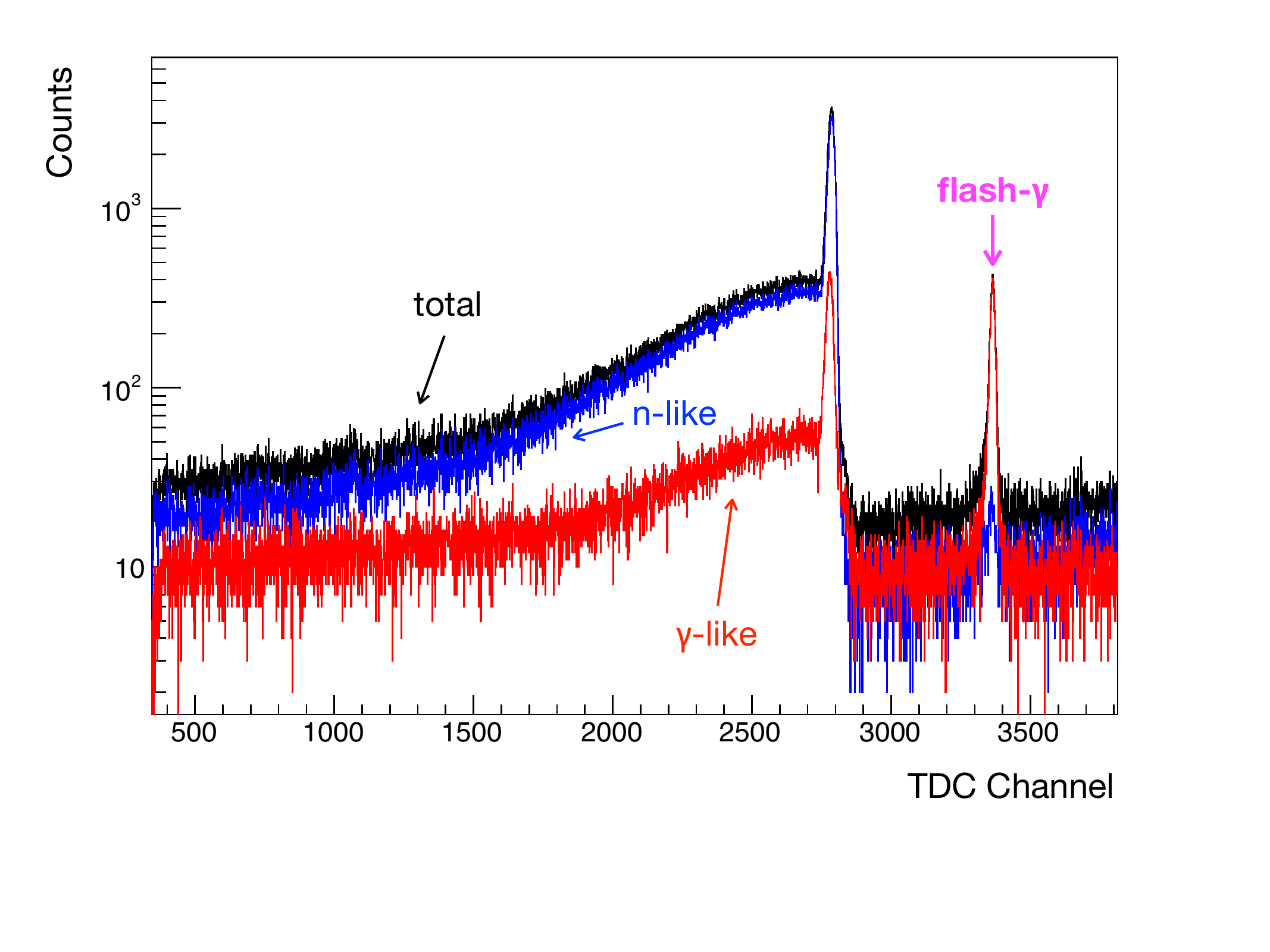}
  \end{center}
  \vspace{-40truept}
  \caption{TOF distributions of all (black), neutron-like (blue), 
           and $\gamma$-like (red) events. 
           The TDC was operated in common stop mode.}
  \label{fig:tofdist}
  \end{figure}

\begin{figure}[htbp]
  \begin{center}
   \includegraphics[clip,width=9.0cm]{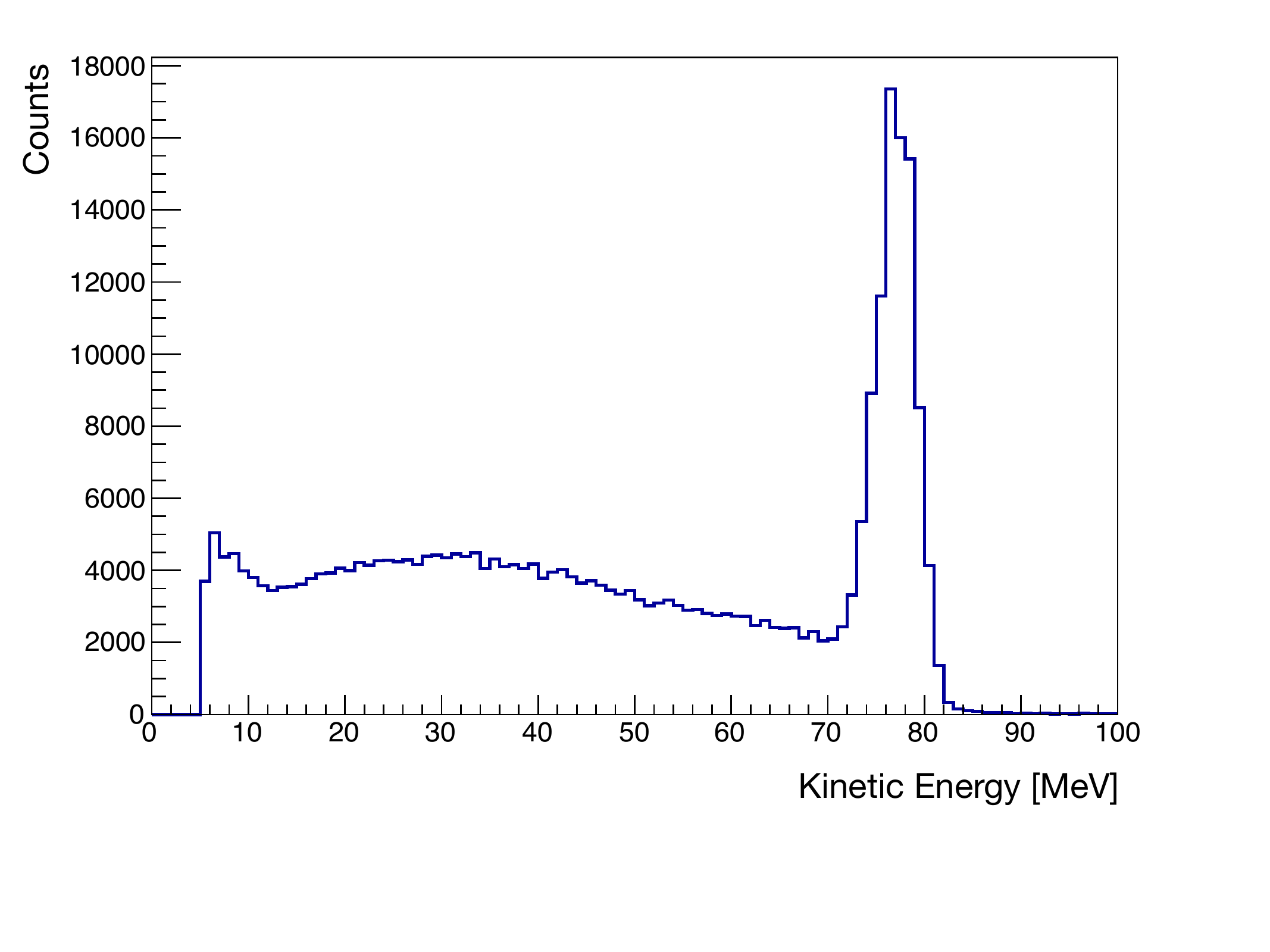}
  \end{center}
  \vspace{-40truept}
  \caption{Neutron kinetic energy distribution reconstructed from 
           the TOF distribution.} 
  \label{fig:lqsekindist}
  \end{figure}

The time distributions of both neutron and $\gamma$-ray candidates are then reconstructed using TDC data.
Time-walk corrections are separately applied for $\gamma$-like and neutron-like events when they are within 
the QDC dynamic range since their pulse shapes differ in general. 
A common factor is used at high energies where the time-walk effect is expected to be negligible.
Figure~\ref{fig:tofdist} shows TOF distributions after applying these corrections.
The sharp peak around TDC channel 3350 corresponds to prompt $\gamma$-rays (called flash $\gamma$-rays) 
emitted from the initial proton-lithium interaction. 
The limited neutron-like contamination in the peak indicates that the PSD cut is functioning well.
Neutron kinetic energies are reconstructed by using the time difference between their interaction 
and the flash $\gamma$-ray peak and the known distance between the lithium target and the scintillator. 
The result is shown in Figure~\ref{fig:lqsekindist}, whose peak at 77~MeV is consistent with 
the expectation from the beamline settings.
The flux measurement below uses more than 50,000 events located in the peak region 
defined by $72 < E_{\rm kin} < 82$~MeV.

\subsection{Neutron detection efficiency}

\begin{figure}[htbp]
  \begin{center}
   \includegraphics[clip,width=9.0cm]{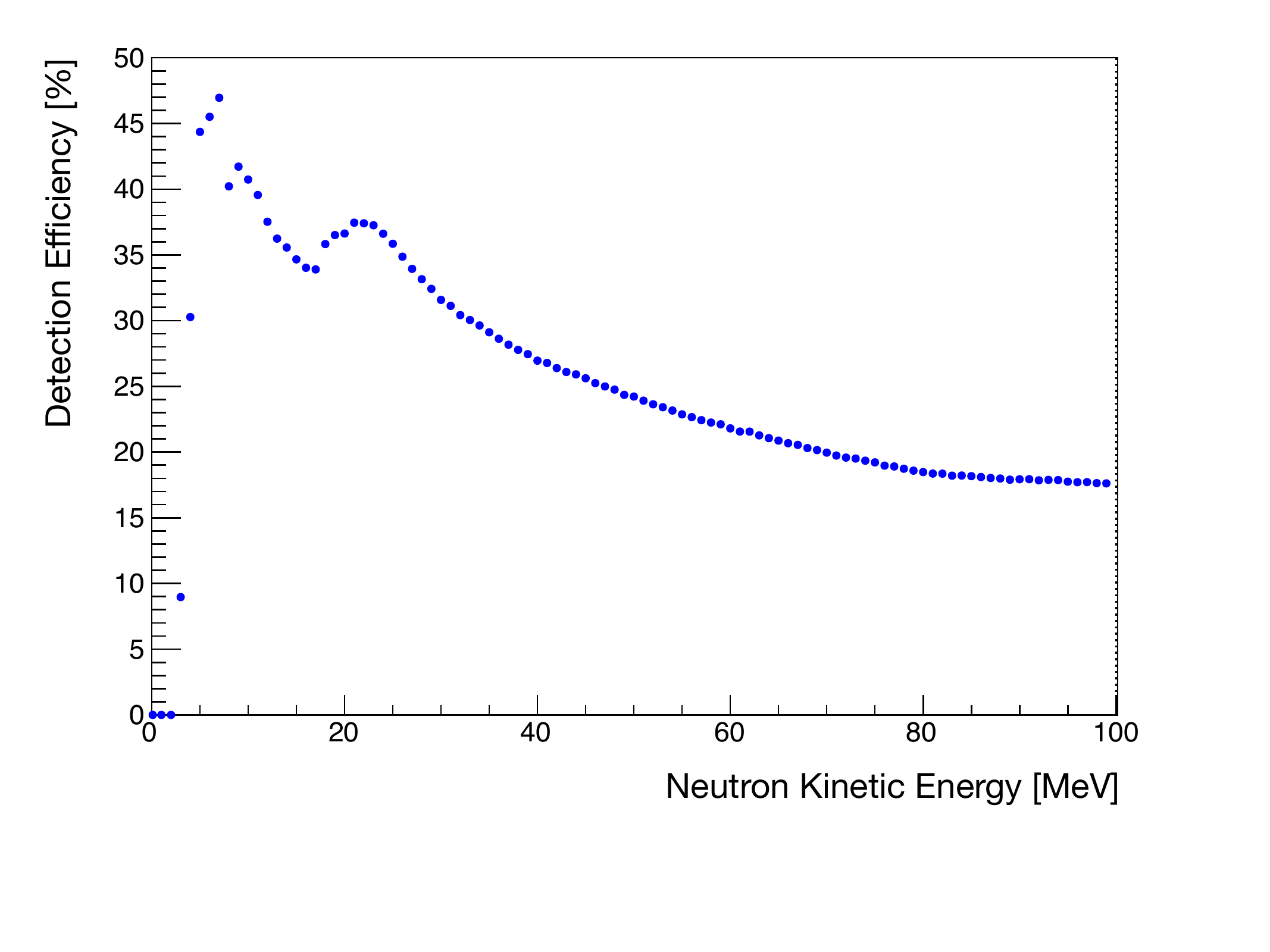}
  \end{center}
  \vspace{-40truept}
  \caption{Neutron detection efficiencies of the BC-501A detector calculated with SCINFUL-QMD.
	   The attenuation factor of 0.008~${\rm cm^{-1}}$ and the light output 
	   function from Ref.~\cite{satoh} are used as the nominal setting.}
  \label{fig:scinfuleff}
  \end{figure}

The neutron detection efficiency of the BC-501A scintillator was calculated 
using the SCINFUL-QMD Monte Carlo (MC) code in each energy bin. 
The inputs to the MC are the detector and source geometries, the detector threshold, 
the light attenuation factor in the BC-501A scintillator, and the PMT's response function.
The detector threshold was obtained using energy calibration data and the scintillator 
attenuation factor, 0.008~${\rm cm^{-1}}$, was adopted from previous measurements~\cite{flux2}. 
SCINFUL-QMD implements three functional forms to describe the PMT light output~\cite{scinful,satoh,nakao}.
The efficiency results with these three functions are compared and their relative 
difference is included as a systematic error in the analysis. 
Here the function from Ref.~\cite{satoh} is used as the nominal setting.
During the simulation 100,000 neutrons are traced in each of 100 energies 
spanning the range 0.1~MeV to 99~MeV, with 1~MeV-wide bins above 1~MeV.
Figure~\ref{fig:scinfuleff} shows the calculated neutron detection efficiencies
with the nominal setting.
The bump seen around 20~MeV is attributed to an open interaction channel on carbon in the BC-501A scintillator. 

\subsection{Flux estimation}

\begin{figure}[htbp]
  \begin{center}
   \includegraphics[clip,width=9.0cm]{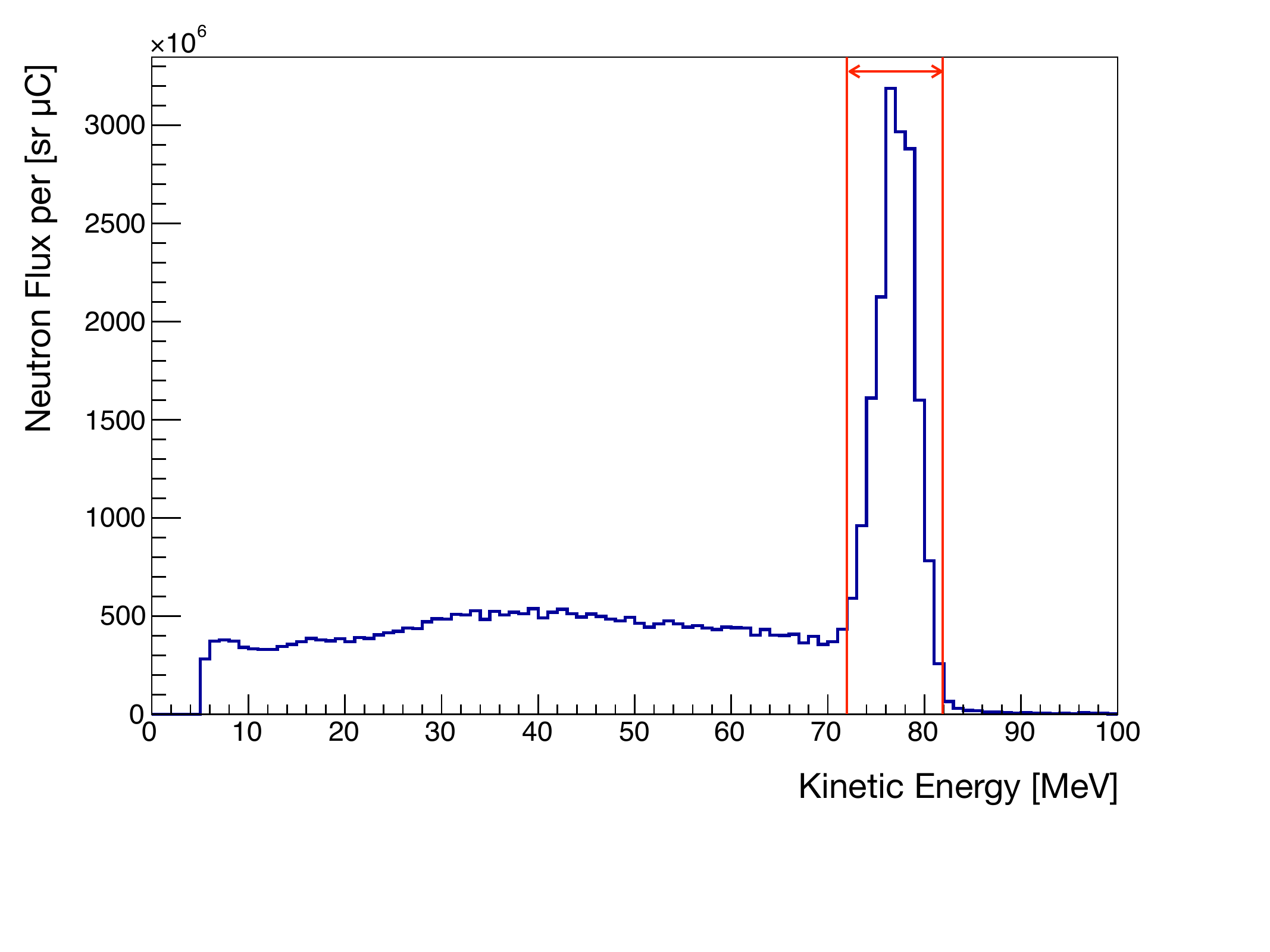}
  \end{center}
  \vspace{-30truept}
  \caption{Neutron flux normalized by the detector covering the solid angle and the incident protons.  
	   The red bars indicate the peak region used for the cross section measurement.}
  \label{fig:fluxdist}
  \end{figure}

The neutron flux is obtained from the kinetic energy distribution corrected by the detector solid 
angle from the lithium target and the detection efficiency in each energy bin and normalized by 
the incident protons.
Figure~\ref{fig:fluxdist} shows the resulting distribution. 
The total flux in the peak region between 72 and 82~MeV is $1.71 \times 10^{10} \ {\rm (sr \ \mu C)^{-1}}$, 
and is consistent with similar measurements using the same beamline~\cite{flux1,flux2}.
Only the peak region is used in the cross section measurement, as below 72~MeV many neutrons have 
scattered before reaching the water sample and are thus considered to be a background.
For the cross section measurement, this flux needs to be modified to account for 
geometric differences between the BC-501A detector and the water sample as well as for 
the difference in the neutron mean free path in each. 
The mean free path for 77~MeV neutrons in water is $\approx$30~cm and is 
$\approx$34~cm in the BC-501A scintillator~\cite{exfor}.
The correction factor, which converts the flux measured by the BC-501A detector 
to the flux in the $\gamma$-ray measurement with water, is defined as: 

  \begin{eqnarray}
   \alpha_{\rm corr} = 
    \frac{\int_{0}^{Z_{\rm Water}} e^{-z/L_{\rm Water}} dz}
         {\int_{0}^{Z_{\rm BC\mathchar`-501A}} e^{-z/L_{\rm BC\mathchar`-501A}} dz} = 1.09, 
  \label{eq:fluxcorr}
  \end{eqnarray}
  \vspace{1truept}

\noindent 
where $z$ denotes the beam direction, and $Z$ and $L$ represent length of the scintillator 
or the water sample and the neutron mean free path of each object, respectively.  
With this correction factor multiplied to the flux measured with the BC-501A, the neutron flux in the $\gamma$-ray 
measurement is obtained to be $\phi_n = 1.87 \times 10^{10} \ {\rm (sr \ \mu C)^{-1}}$. 

\subsection{Flux uncertainties}

This section details the uncertainty estimates in the flux measurement.
The statistical error of the data is less than 0.5\% for the peak region (72$-$82~MeV). 
Table~\ref{tab:fluxsys} summarizes the statistical and systematic errors. 

  \begin{table}[htbp]
  \begin{center}
  \caption{Statistical and systematic uncertainties of the neutron flux measurement.}
  \label{tab:fluxsys}
  \vspace{2truept}
   \begin{tabular}{l c}                                  \hline \hline
    Error source                     & Size [\%]   \\ \hline
    Statistical                      & 0.5         \\ \hline
    Beam stability                   & 1.4         \\
    Neutron selection                & 2.2         \\
    Detection efficiency by SCINFUL-QMD   & 10.0   \\
    Kinetic energy reconstruction    & 1.0         \\ 
    Former bunch and environmental events & 1.0    \\
    Correction from BC-501A to water & 3.7         \\ \hline 
    Total                            & 11.2        \\ \hline \hline 
   \end{tabular}
  \end{center}
  \end{table}

\subsubsection{Beam stability}

The neutron flux was measured at the beginning, the middle, and the end of the experiment. 
Figures~\ref{fig:psddist} to \ref{fig:fluxdist} show the results from the final measurement.
Over the three measurements the flux was stable within 1.4\%. 
The average flux is used for the cross section measurement and this variation is 
incorporated as a systematic error.

\subsubsection{Neutron selection}

As described above, the PSD cut is used to extract neutrons with energies within the range of the QDC. 
The uncertainty of this cut is estimated using the contamination of neutron-like events 
in the flash $\gamma$-ray peak in Figure~\ref{fig:tofdist}.
This results in a 2.0\% uncertainty in the neutron flux. 
In addition, the contamination of $\gamma$-ray events in the higher energy data is extrapolated 
into the QDC overflow region from Figure~\ref{fig:lqsspectrum}. 
This yields a contamination of 0.8\%.
Accordingly, the neutron selection error is taken to be the sum in quadrature of these two 
components, 2.2\% in total.

\subsubsection{Detection efficiency by SCINFUL-QMD}

The uncertainty related to the physics model of SCINFUL-QMD is estimated to be 10\% for energies 
below 80~MeV based on previous studies~\cite{scinfulqmd1,satoh,flux2}.
The MC statistical error is 0.3\%. 
The systematic error related to the threshold value coming from the energy calibration error is 
estimated to be less than 0.1\%.
Conservatively adjusting the light attenuation factor in the simulation was found to have 
a negligible effect on the efficiency. 
Similarly, the selection of the light output function does not produce more than a 0.1\% change 
in the result.
In total, a 10.0\% uncertainty is assigned to the efficiency calculation.

\subsubsection{Kinetic energy reconstruction and contributions from previous bunches}

Systematic errors in the timing measurement can result in uncertainties in the reconstructed 
kinematic energy and subsequently flux due to efficiency differences between energy bins.
While the time-walk correction was found to have negligible impact on the analysis, 
the calibration of the TDC leads to a 0.4~ns uncertainty in the TOF measurement.
Alignment uncertainties in the detector setup produce a 0.3~ns error and the width of 
flash $\gamma$-ray peak incurs a further 1.1~ns. 
In total a 1.2~ns uncertainty is assigned to the TOF measurement, which corresponds to a 1~MeV 
uncertainty in the kinetic energy reconstruction and a less than 1\% error in the flux estimation. 
 
Contamination from the prior beam bunches and environmental neutrons was estimated by comparing 
the event rate in the region between the flash $\gamma$-ray and the neutron peaks to that in 
the neutron peak region in Figure~\ref{fig:tofdist}.
The contamination amount is found to be less than 1\%.

\subsubsection{Flux correction}

The correction factor ($\alpha_{\rm corr}$) is affected by uncertainties on 
the neutron reaction cross section and geometry. 
Since the correction factor is made from the ratio of the scintillator and the water sample, 
these model uncertainties nearly cancel, leaving a remaining uncertainty of 3.7\%. 
The effect of the geometrical error on the correction factor is negligible.

\subsection{Neutron beam profile}

\begin{figure}[htbp]
  \begin{center}
   \includegraphics[clip,width=9.0cm]{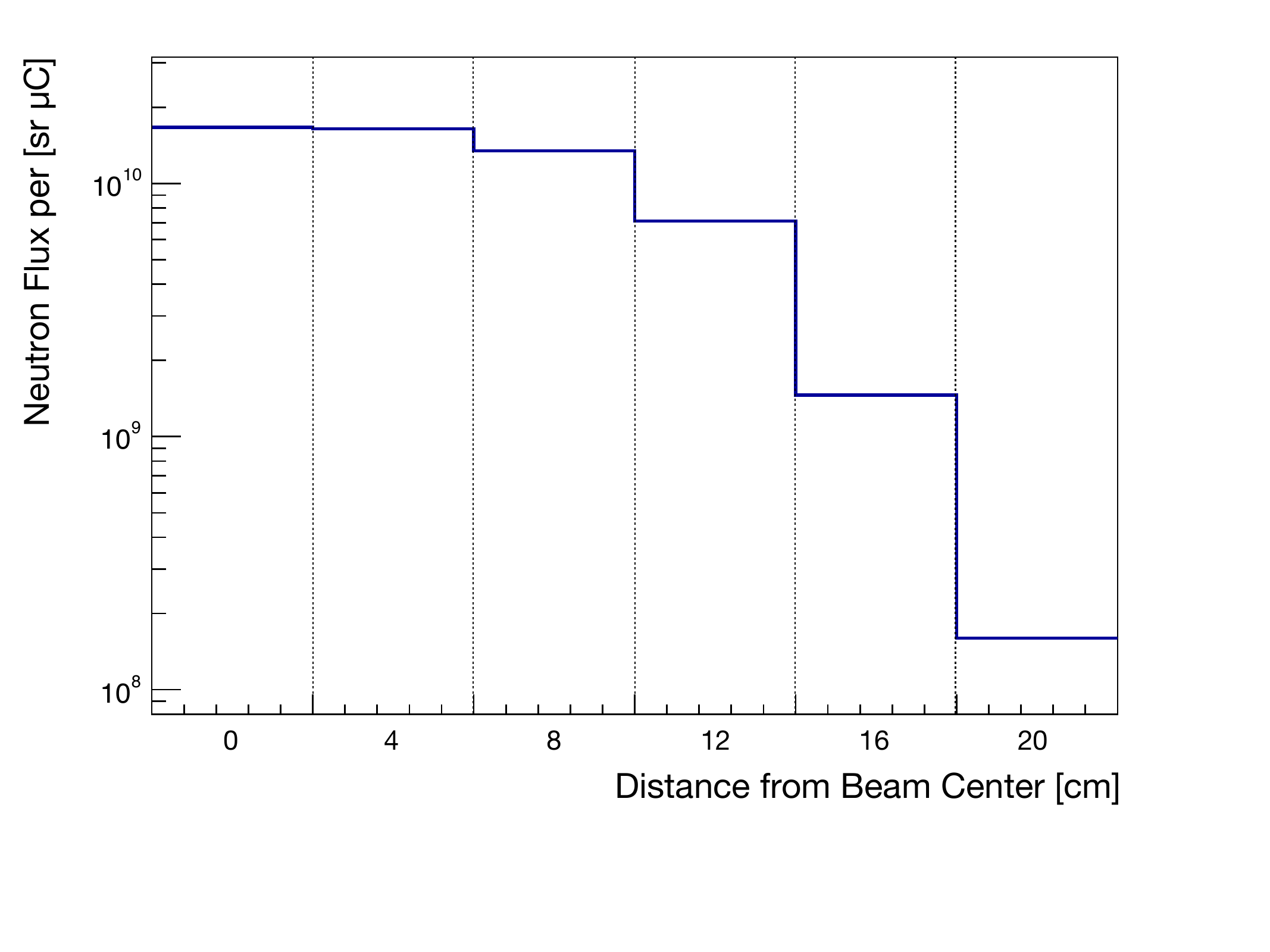}
  \end{center}
  \vspace{-40truept}
  \caption{Neutron beam profile measured by the BC-501A detector.
           The results are for the 72$-$82~MeV region.}
  \label{fig:fluxprofile}
  \end{figure}

In order to reduce neutron backgrounds in the $\gamma$-ray detectors resulting from direct 
exposure to the beam, a profile scan was conducted ahead of the $\gamma$-ray measurements.
During the scan the BC-501A scintillator's center was shifted from directly on the beam 
axis to 20~cm perpendicularly off-axis in steps of 4~cm.
The flux was measured using the same method as described above and the result for the peak region 
($72 < E_{\rm kin} < 82$~MeV) appears in Figure~\ref{fig:fluxprofile}.
The neutron flux 20~cm away from the beam center is smaller than that at the center
by more than two orders of magnitude.
Further, since this is outside the expected beam profile as determined by the collimator 
(10~cm from the beam center), the $\gamma$-ray detectors were placed in this position. 
Neutron backgrounds at this position were measured with the CsI(Tl) scintillator
as explained in Section~\ref{sec:gammaray}.

\subsection{TOF measurement in ${\bf LaBr_3(Ce)}$}

\begin{figure}[htbp]
  \begin{center}
   \includegraphics[clip,width=9.0cm]{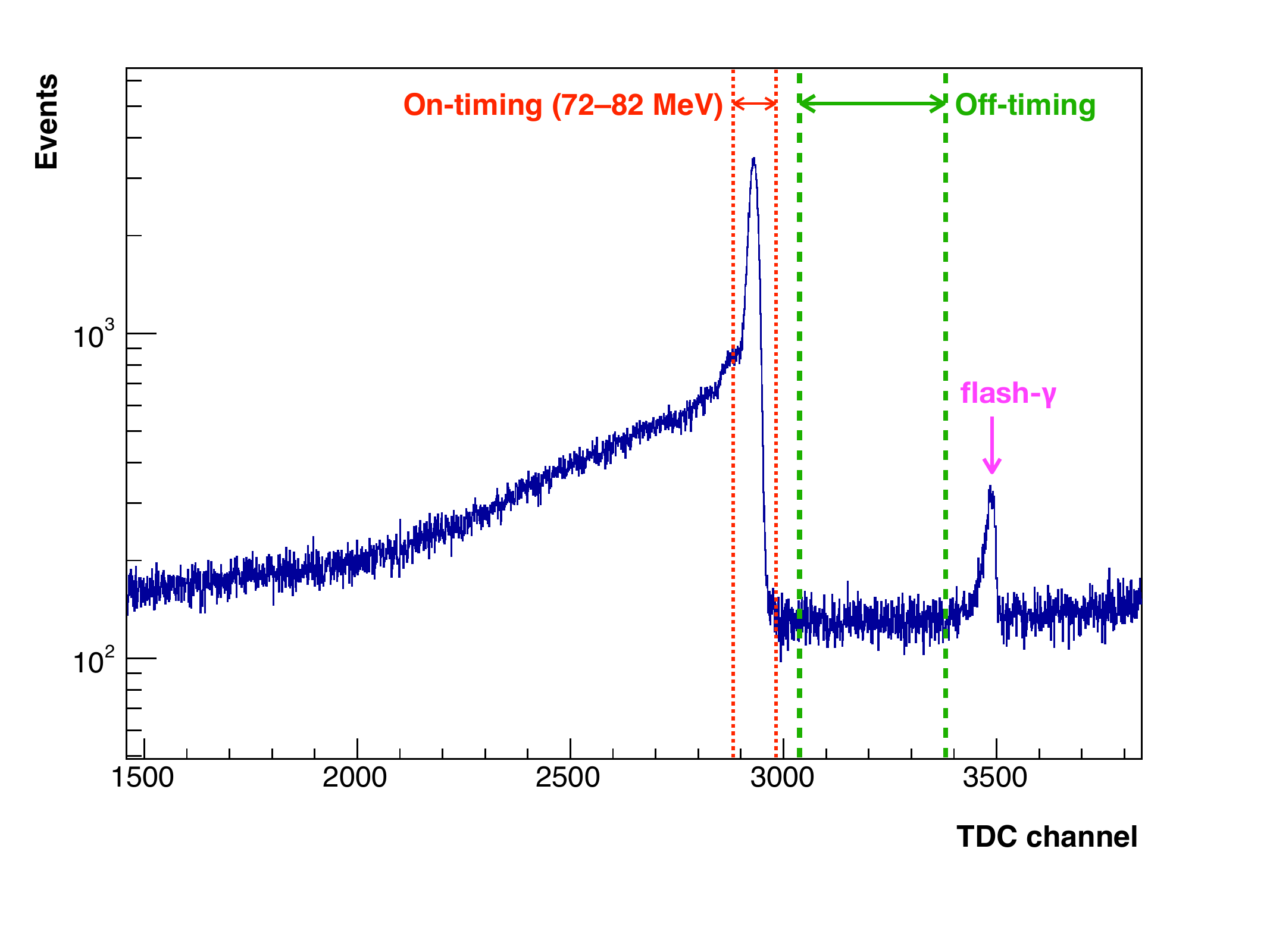}
  \end{center}
  \vspace{-25truept}
  \caption{TOF distribution of observed events in the ${\rm LaBr_3(Ce)}$ detector.
	   The red and green bars indicate the ``on-timing" and ``off-timing" regions, respectively.
           The TDC was operated in common stop mode.}
  \label{fig:labrtofdist}
  \end{figure}

\begin{figure}[htbp]
  \begin{center}
   \includegraphics[clip,width=9.0cm]{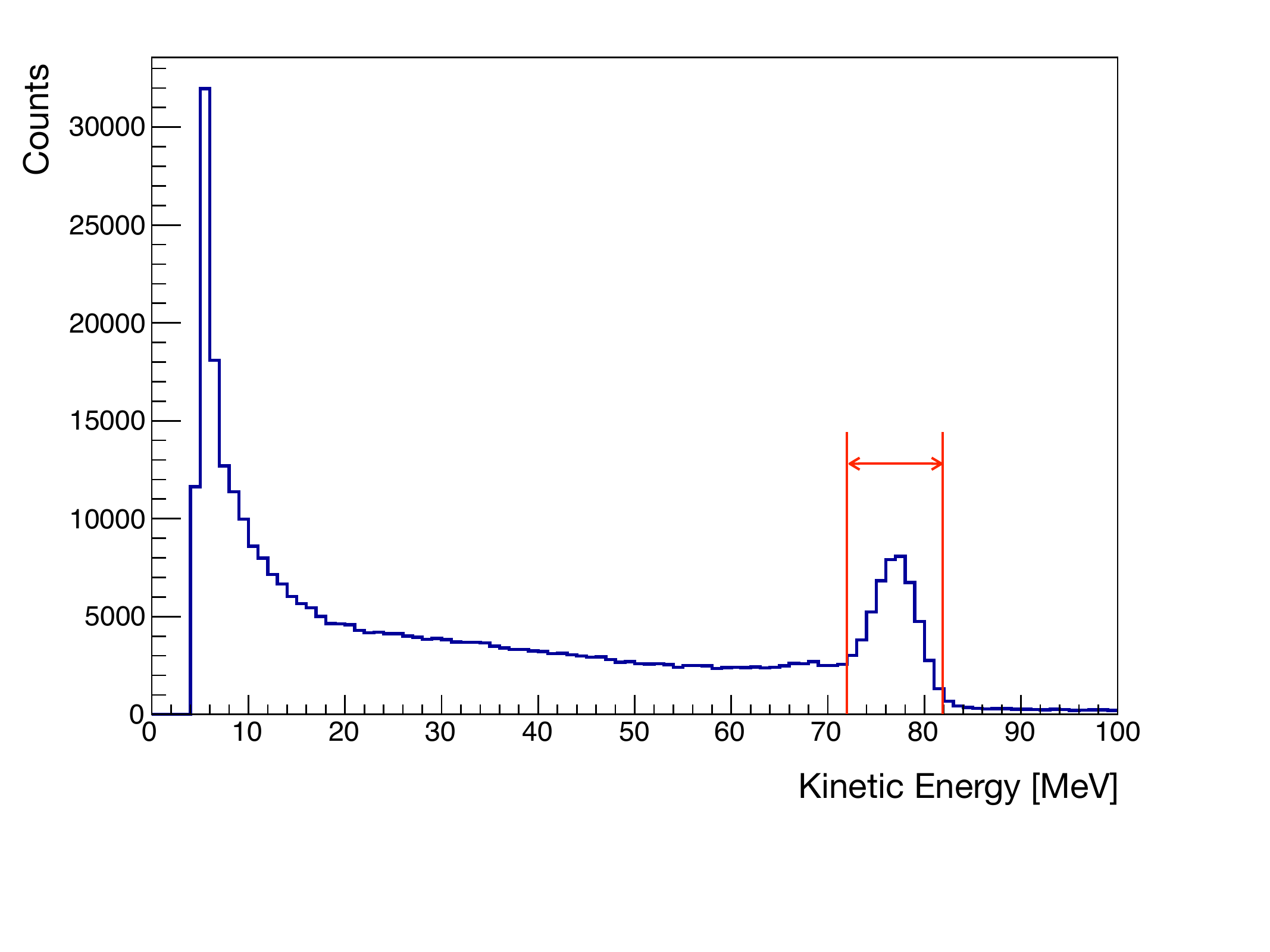}
  \end{center}
  \vspace{-35truept}
  \caption{Neutron kinetic energy distribution inferred from timing information 
           of observed events in the ${\rm LaBr_3(Ce)}$ detector.
	   The red bars indicate the peak region used for the cross section measurement.} 
  \label{fig:labrekindist}
  \end{figure}

To infer the kinetic energies of neutrons producing $\gamma$-rays observed 
in the ${\rm LaBr_3(Ce)}$ detector, timing information is used to perform 
a TOF analysis similar to that for the BC-501A detector. 
The $\gamma$-ray event timing is corrected for time-walk effects and the distance 
between the detector and the acrylic vessel when reconstructing these kinetic energies. 
Figure~\ref{fig:labrtofdist} shows the TOF distribution from the ${\rm LaBr_3(Ce)}$ detector, 
and Figure~\ref{fig:labrekindist} shows the distribution of inferred neutron kinetic energies.
The neutron flux peak position and width are consistent with the measurement from the BC-501A scintillator.
For later analysis, the ``on-timing" and ``off-timing" regions are defined 
as shown in Figure~\ref{fig:labrtofdist}. 
The on-timing region corresponds to events whose reconstructed kinetic energy 
is between 72 and 82~MeV. 

\section{Gamma-ray production}
\label{sec:gammaray}

\begin{figure*}[htbp]
  \begin{center}
   \includegraphics[clip,width=17.0cm]{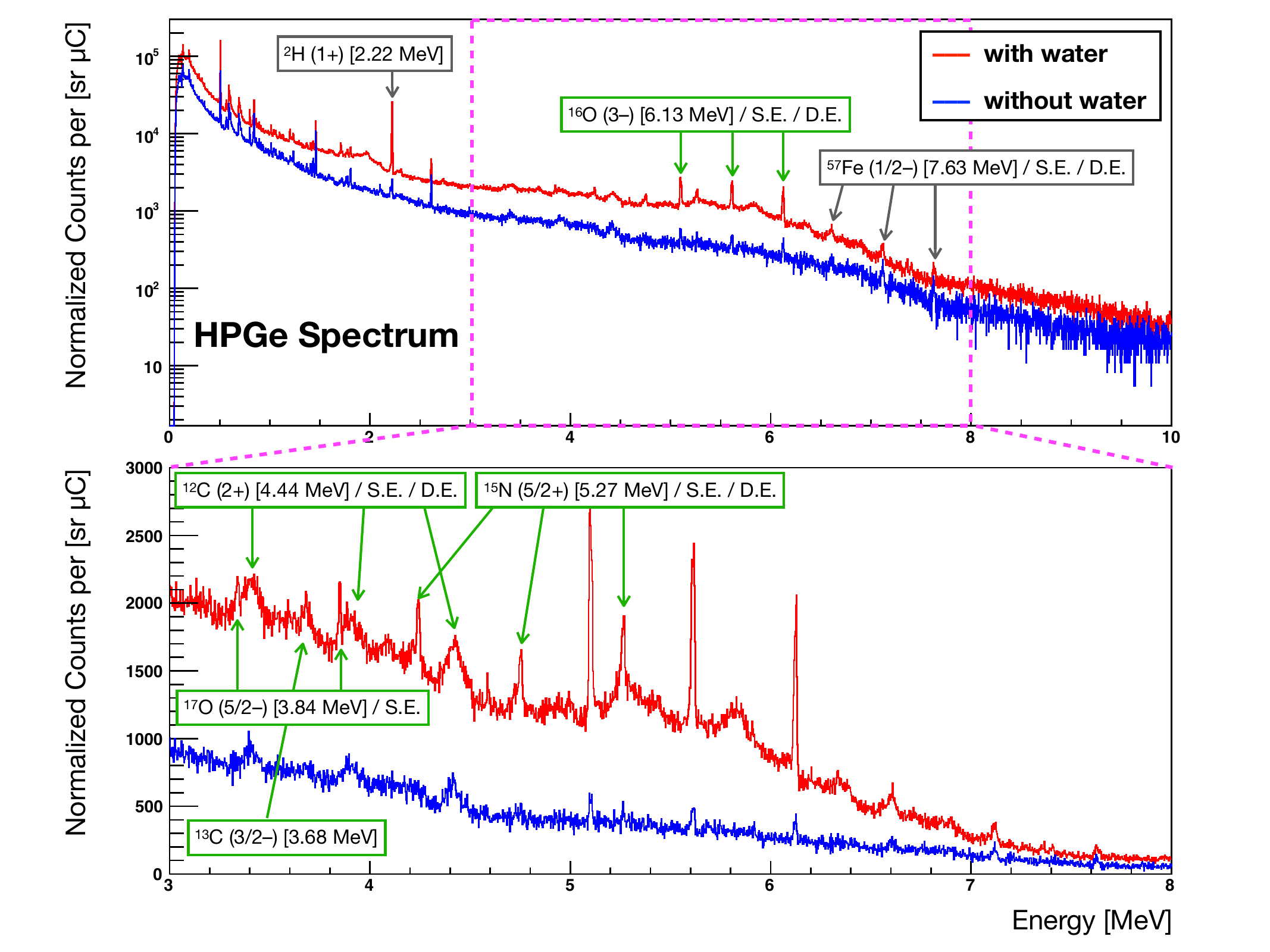}
  \end{center}
  \vspace{-10truept}
  \caption{Energy spectra of the HPGe detector with water (red) and without water (blue).
           The bottom panel gives the region between 3 and 8~MeV.}
  \label{fig:hpgespectrum}
  \end{figure*}

\begin{figure*}[htbp]
  \begin{center}
   \includegraphics[clip,width=17.0cm]{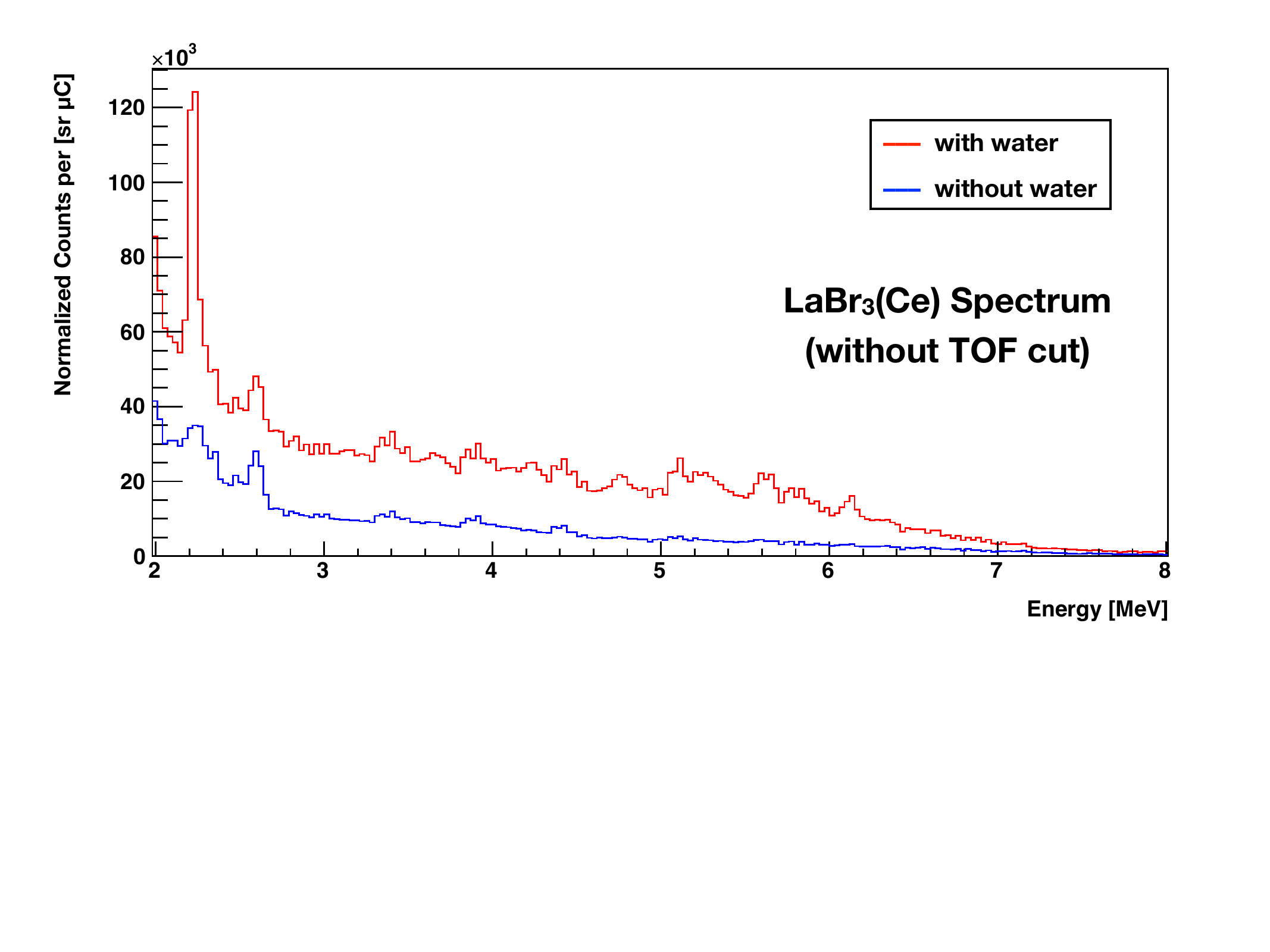}
  \end{center}
  \vspace{-130truept}
  \caption{Energy spectra of the ${\rm LaBr_3(Ce)}$ detector with water (red) and without 
           water (blue) before the TOF cut.}
  \label{fig:labrspectrum_notof}
  \end{figure*}

\begin{figure*}[htbp]
  \begin{center}
   \includegraphics[clip,width=17.0cm]{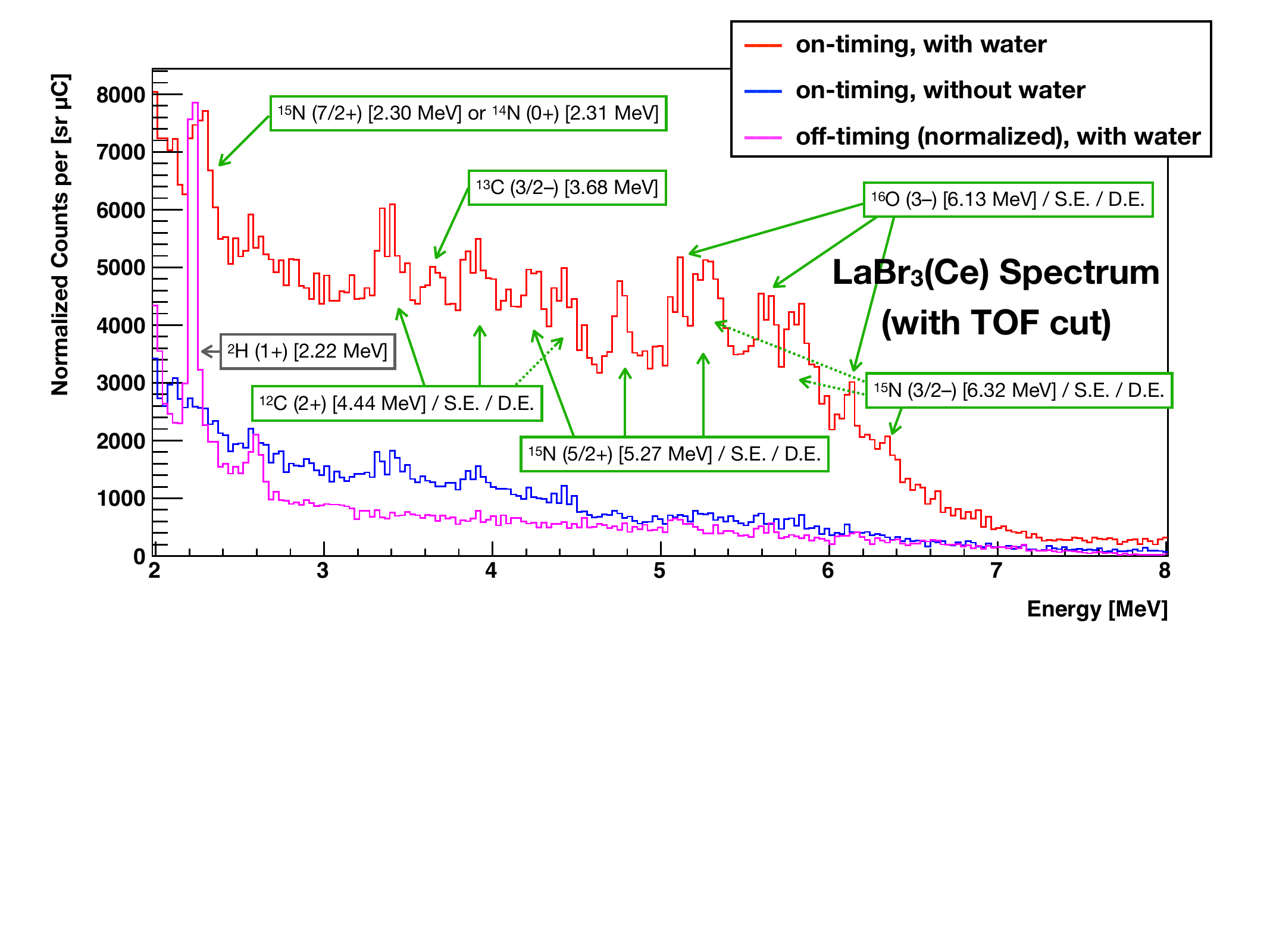}
  \end{center}
  \vspace{-130truept}
  \caption{Energy spectra of the ${\rm LaBr_3(Ce)}$ detector with three different conditions:
           on-timing with water, on-timing without water, and off-timing with water.}
  \label{fig:labrspectrum_tof}
  \end{figure*}

Figure~\ref{fig:hpgespectrum} shows the observed energy spectrum measured with the HPGe detector 
and Figure~\ref{fig:labrspectrum_notof} shows the ${\rm LaBr_3(Ce)}$ spectrum without 
the TOF cut. 
The red and blue spectra are the results with water-filled and empty 
configurations, respectively. 
The spectra are normalized using the solid angle covered by the acrylic container as viewed from 
the lithium target and the incident proton beam. 
The total injected protons on the lithium target are 1.65 (1.27) mC for 
the water-filled (empty) configuration.
Several $\gamma$-ray peaks are observed in both detectors. 
The ${\rm LaBr_3(Ce)}$ spectra with the TOF cut are presented in Figure~\ref{fig:labrspectrum_tof}.
In this figure, three spectra with different conditions are shown: 
one in a water-filled setup and with the on-timing cut, 
one in an empty setup and with the on-timing cut, and 
one in a water-filled setup and with the off-timing cut.
The spectrum with the off-timing cut is normalized to the length of the on-timing window 
defined in Figure~\ref{fig:labrtofdist}.
The $\gamma$-ray peaks of primary interest to the present measurement, their parent nuclei and 
excited states, and the physics processes which produce them are summarized in Table~\ref{tab:gammapeaks}. 
Parent nuclei are identified by the energy and width of their peaks.
Some peaks are from nuclei with shorter decay times than the duration of the recoil they 
incur after being struck by an incident particle. 
This produces a Doppler effect on the resulting $\gamma$-ray, which broadens the observed peak. 

  \begin{table*}[htbp]
  \begin{center}
  \caption{Summary on $\gamma$-ray energies, parent nuclei with their spin ($J$) and 
           parity ($\pi$), decay lifetime, and physics processes.}
  \label{tab:gammapeaks}
  \vspace{2truept}
   \begin{tabular}{c c c c} \hline \hline
    Energy [MeV] \ & Parent ($J^{\pi}$) \ & $T_{1/2}$ \cite{endf} & Physics process \\ \hline
    7.12 \ & ${\rm ^{16}O(1^{-})}$           \ & 8.3~fs \ &
     ${\rm ^{16}O}(n,n'){\rm ^{16}O^{*}}$ \\
    6.92 \ & ${\rm ^{16}O(2^{+})}$           \ & 4.70~fs \ &
     ${\rm ^{16}O}(n,n'){\rm ^{16}O^{*}}$ \\
    6.32 \ & ${\rm ^{15}N(\frac{3}{2}^{-})}$ \ & 0.146~fs \ &
     ${\rm ^{16}O}(n,np){\rm ^{15}N^{*}}$ \\
    6.13 \ & ${\rm ^{16}O(3^{-})}$           \ & 18.4~ps \ &
     ${\rm ^{16}O}(n,n'){\rm ^{16}O^{*}}$ \\
    5.27 \ & ${\rm ^{15}N(\frac{5}{2}^{+})}$ \ & 1.79~ps \ &
     ${\rm ^{16}O}(n,n'){\rm ^{16}O^{*}}$ then ${\rm ^{16}O^{*}} \rightarrow {\rm ^{15}N^{*}} + p$,
     or ${\rm ^{16}O}(n,np){\rm ^{15}N^{*}}$ \\
    5.18 \ & ${\rm ^{15}O(\frac{1}{2}^{+})}$ \ & 5.7~fs \ &
     ${\rm ^{16}O}(n,n'){\rm ^{16}O^{*}}$ then ${\rm ^{16}O^{*}} \rightarrow {\rm ^{15}O^{*}} + n$,
     or ${\rm ^{16}O}(n,2n){\rm ^{15}O^{*}}$ \\
    4.44 \ & ${\rm ^{12}C(2^{+})}$           \ & 60.9~fs \ &
     ${\rm ^{16}O}(n,n'){\rm ^{16}O^{*}}$ then ${\rm ^{16}O^{*}} \rightarrow {\rm ^{12}C^{*}} + \alpha$,
     or ${\rm ^{16}O}(n,n\alpha){\rm ^{12}C^{*}}$ \\
    3.68 \ & ${\rm ^{13}C(\frac{3}{2}^{-})}$ \ & 1.10~fs \ &
     ${\rm ^{16}O}(n,\alpha){\rm ^{13}C^{*}}$ \\
    2.31 \ & ${\rm ^{14}N(0^{+})}$           \ & 68~fs \ &
     ${\rm ^{16}O}(n,2np){\rm ^{14}N^{*}}$ \\
    2.30 \ & ${\rm ^{15}N(\frac{7}{2}^{+})}$ \ & 8~fs \ &
     ${\rm ^{16}O}(n,np){\rm ^{15}N^{*}}$ \\ \hline \hline
   \end{tabular}
  \end{center}
  \end{table*}

The 6.13~MeV $\gamma$-ray from the excited state of ${\rm ^{16}O}$ is clearly observed
by both the HPGe and ${\rm LaBr_3(Ce)}$ detectors.
This is expected to be produced by the $(n,n')$ inelastic scattering. 
The peak appears stronger in the spectra without the TOF cut, as shown in 
Figures~\ref{fig:hpgespectrum} and \ref{fig:labrspectrum_notof}, than with the TOF cut, 
as shown in the red spectrum in Figure~\ref{fig:labrspectrum_tof}. 
This may be due to large contributions from the lower energy neutrons.
%
The 6.92 and 7.12~MeV $\gamma$-rays, which are emitted after the $(n,n')$ scattering,
are more probable above 6.5~MeV, because these two are from one and two higher 
excited states than 6.13~MeV, respectively.
Since there are seen some contributions considered to be from the neutron-oxygen reaction above 
6.5~MeV, these two components are considered in the spectrum fitting as explained later.

A large bump around 5.8~MeV in the ${\rm LaBr_3(Ce)}$ spectrum with the on-timing cut 
(Figure~\ref{fig:labrspectrum_tof}) is hard to explain by only the Compton edge of 
the 6.13~MeV $\gamma$-ray peak.
It is instead thought to arise from the 6.32~MeV $\gamma$-ray from ${\rm ^{15}N}$. 
This peak is considered to come from the direct knock-out process, $(n,np)$, because
the 6.32~MeV $\gamma$-ray emission is dominant when ${\rm ^{15}N}$ is created via 
$(n,np)$ according to Refs.~\cite{ejiri,leuschner}.
This peak is not observed clearly in the spectra without the TOF cut in
Figures~\ref{fig:hpgespectrum} and \ref{fig:labrspectrum_notof}, because contributions 
from other interactions by lower energy neutrons, which are likely to produce the 6.13~MeV 
$\gamma$-ray, may be dominant. 
There is a similar direct knock-out process $(n,2n)$; however, the 6.18~MeV $\gamma$-rays 
from the excited state after this process are not observed clearly in this experiment.
This may be because neutrons are more likely to be paired with protons inside nuclei, 
therefore the $(n,np)$ process is more probable to occur than the $(n,2n)$ process.

The 5.27~MeV $\gamma$-ray from ${\rm ^{15}N(\frac{5}{2}^+)}$ is clearly seen in 
the HPGe spectrum (Figure~\ref{fig:hpgespectrum}). 
This peak is less visible in the ${\rm LaBr_3(Ce)}$ spectrum without the TOF cut 
(Figure~\ref{fig:labrspectrum_notof}) because of its poorer resolution compared to the HPGe. 
However, with the on-timing TOF cut, contributions from this $\gamma$-ray are 
visible especially around the second escape (S.E.) position in Figure~\ref{fig:labrspectrum_tof}. 
Possible physics processes which produce the 5.27~MeV $\gamma$-ray are nucleon knock-out, 
${\rm ^{16}O}(n,np)$, deuteron flipping, ${\rm ^{16}O}(n,d)$, and nuclear decay from an excited 
state of ${\rm ^{16}O}$ with proton emission, ${\rm ^{16}O^{*} \rightarrow ^{15}N^{*}} + p$, 
after the $(n,n')$ scattering. 
In the present work, these processes are not distinguished, hence an inclusive measurement 
is performed.
It is worth noting that the ${\rm ^{16}O}(n,np)$ cross section prediction is small 
at a neutron energy of 60.7~MeV in Ref.~\cite{dimbylow}  and the 6.32~MeV $\gamma$-ray 
is the most likely if the direct knock-out process occurs \cite{ejiri,leuschner}. 
Therefore, the 5.27~MeV $\gamma$-ray here may originate from the $(n,n')$ scattering 
followed by nuclear decay with proton emission.   
%
The 4.44~MeV $\gamma$-ray from ${\rm ^{12}C(2^+)}$ is also observed.
Here alpha knock-out, ${\rm ^{16}O}(n,n\alpha)$, and decay of ${\rm ^{16}O}$ with alpha 
emission (${\rm ^{16}O^{*} \rightarrow ^{12}C^{*} + \alpha}$) (c.f.~Ref.~\cite{dimbylow})
are potential physics processes that can produce the 4.44~MeV $\gamma$-ray.
Similarly to the case for the 5.27~MeV peak, these processes are not distinguished in 
the analysis and then an inclusive measurement is performed. 
%
Similarly the 5.18~MeV $\gamma$-ray from ${\rm ^{15}O(\frac{1}{2}^+)}$ with subsequent neutron emission 
is expected but is not observed clearly in the present experiment. 
This may be understood by the fact that the minimum excited energy required for nuclear decay 
with neutron emission, 15.66~MeV, is higher than those for nuclear decay with proton emission, 
12.13~MeV, and alpha emission, 7.16~MeV. 
The 5.18~MeV $\gamma$-ray is, however, considered in the spectrum analysis. 
Indeed, inclusion of this peak gives a better fit to the data. 

The 3.68~MeV $\gamma$-ray is observed in both the HPGe and ${\rm LaBr_3(Ce)}$ spectra.
This $\gamma$-ray is considered to be emitted from ${\rm ^{13}C}$ generated by ${\rm ^{16}O}(n,\alpha)$ reactions.
Another peak is observed clearly around 2.30~MeV, which is not as visible in the spectra 
without the TOF cut.
It is obscured by the intense peak at 2.22~MeV $\gamma$-ray, which is produced from 
thermal neutron capture on hydrogen.
The thermal neutron induced events can be removed using off-timing data, as explained 
in the next section.
There are two possibilities for the 2.30~MeV peak: the 2.30~MeV $\gamma$-ray from ${\rm ^{15}N(\frac{7}{2}^{+})}$
and the 2.31~MeV $\gamma$-ray from ${\rm ^{14}N(0^{+})}$. 
These two cannot be distinguished by the ${\rm LaBr_3(Ce)}$ due to insufficient energy resolution.

Many peaks which do not originate from fast neutron reactions on oxygen are also observed.
Though they are explained in the following, they are not the main interest of the present work.
The 3.84~MeV $\gamma$-ray seen in Figure~\ref{fig:hpgespectrum} from ${\rm ^{17}O}$ is
thought to come from thermal neutron capture on ${\rm ^{16}O}$.
The 2.22~MeV and 7.63~MeV $\gamma$-rays are likely due to neutron capture on 
${\rm ^{1}H}$ and ${\rm ^{56}Fe}$, respectively. 
Other peaks such as the 1.46~MeV $\gamma$-ray from ${\rm ^{40}K}$ and the 2.61~MeV $\gamma$-ray
from ${\rm ^{208}Tl}$ can be made by a number of neutron reactions with materials in the beamline. 

In this paper, production cross sections for the ten $\gamma$-rays in Table~\ref{tab:gammapeaks}
are measured with a spectrum analysis of the on-timing ${\rm LaBr_3(Ce)}$ data 
(Figure~\ref{fig:labrspectrum_tof}), as explained in Section~\ref{sec:xsection}.
An inclusive cross section is measured for the 2.30~MeV and 2.31~MeV peaks.

\section{Background Estimation}
\label{sec:bkgestimate}

Backgrounds are categorized into four types:
(1) fast neutron reactions with the detector, (2) non-water background, 
(3) $\gamma$-rays from thermal neutron capture and $\beta$'s from beta decay, 
and (4) $\gamma$-rays from scattered fast neutron reactions. 
Each of them is explained in the following.

\subsection{Fast neutron reactions with the detector}

\begin{figure}[htbp]
  \begin{center}
   \includegraphics[clip,width=9.0cm]{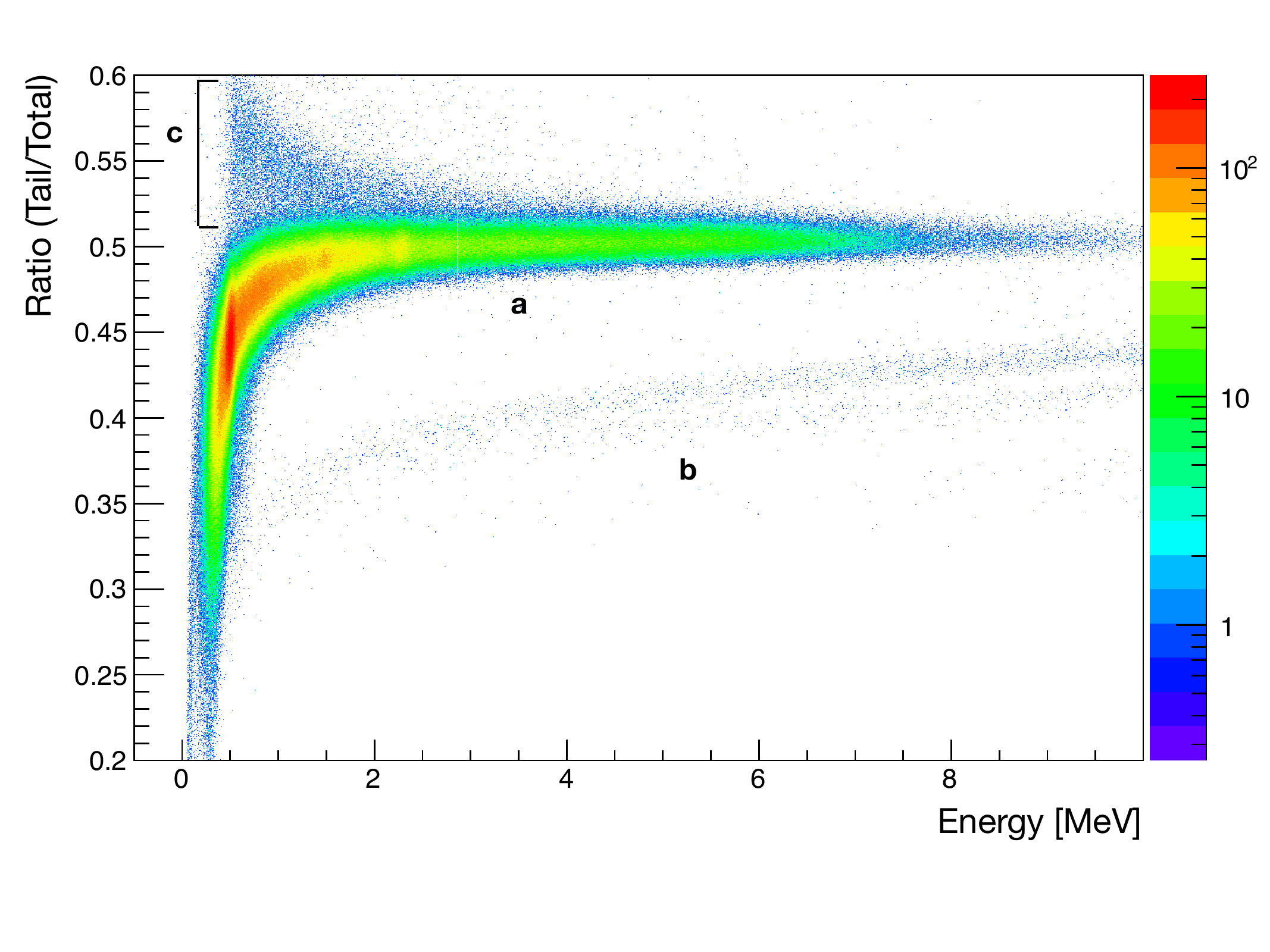}
  \end{center}
  \vspace{-30truept}
  \caption{Distribution of the ratio of the integrated tail to total signal pulse
           of the CsI(Tl) pulse height as a function of deposited energy. 
	   Three populations are seen: (a) $\gamma$-rays, (b) neutrons, and (c) pile-up events.}
  \label{fig:csipsd}
  \end{figure}

Neutrons reacting with the ${\rm LaBr_3(Ce)}$ detector are a potential background 
since they either scatter off the acrylic container or are not on the beam axis.
The CsI(Tl) scintillator was used to measure this background with its PSD capability 
in a manner analogous to that with the BC-501A scintillator.
For this measurement the charge integration region was optimized with a figure-of-merit
laid out in Ref.~\cite{csipsd}.
The same neutron energy region as the cross section analysis, 72$-$82~MeV, is selected
using the TOF cut. 
The ratio of the integrated tail-to-total signal pulse as a function of the deposited energy 
is shown in Figure~\ref{fig:csipsd}.
Here three populations are seen: (a) $\gamma$-rays, (b) neutrons, and (c) pile-up events.
Population (c) is due to events having multiple signals within one Flash-ADC time window.
The number of such pile-up events is negligible compared to the number of $\gamma$-ray events.
The fast neutron background is estimated in each deposited energy region
by subtracting the number of events without water from that with water. 
The resulting contamination of fast neutron reactions is smaller than 1\% in 
every deposited energy region.
This background is found to be negligible compared to the total systematic error 
in the measurement.
Even if the material difference between CsI(Tl) and ${\rm LaBr_3(Ce)}$ is taken into account,
the rate of the fast neutron contamination is still negligible.

\subsection{Non-water background}

Backgrounds originating from neutron reactions on materials other than water are estimated 
using data with the empty vessel.
This is shown as the blue spectrum in Figure~\ref{fig:labrspectrum_tof} and subtracted from 
the result with water (the red spectrum in the same figure).

\subsection{$\gamma$-rays from thermal neutron capture and $\beta$'s from beta decay}

The $\gamma$-rays from thermal neutron capture and the electrons or positrons
from beta decay occur at much longer time scales than the beam repetition cycle, $\approx$560~ns, 
and hence are expected to be distributed uniformly in time. 
Contributions from these backgrounds can be estimated using the off-timing region of the spectrum 
as shown in Figure~\ref{fig:labrspectrum_tof}.
The energy spectrum with the off-timing cut is subtracted from that with the on-timing cut applying 
a normalization based on the length of the time window in Figure~\ref{fig:labrtofdist}.

\subsection{$\gamma$-rays from scattered fast neutron reactions}

\begin{figure*}[htbp]
  \begin{center}
   \includegraphics[clip,width=16.0cm]{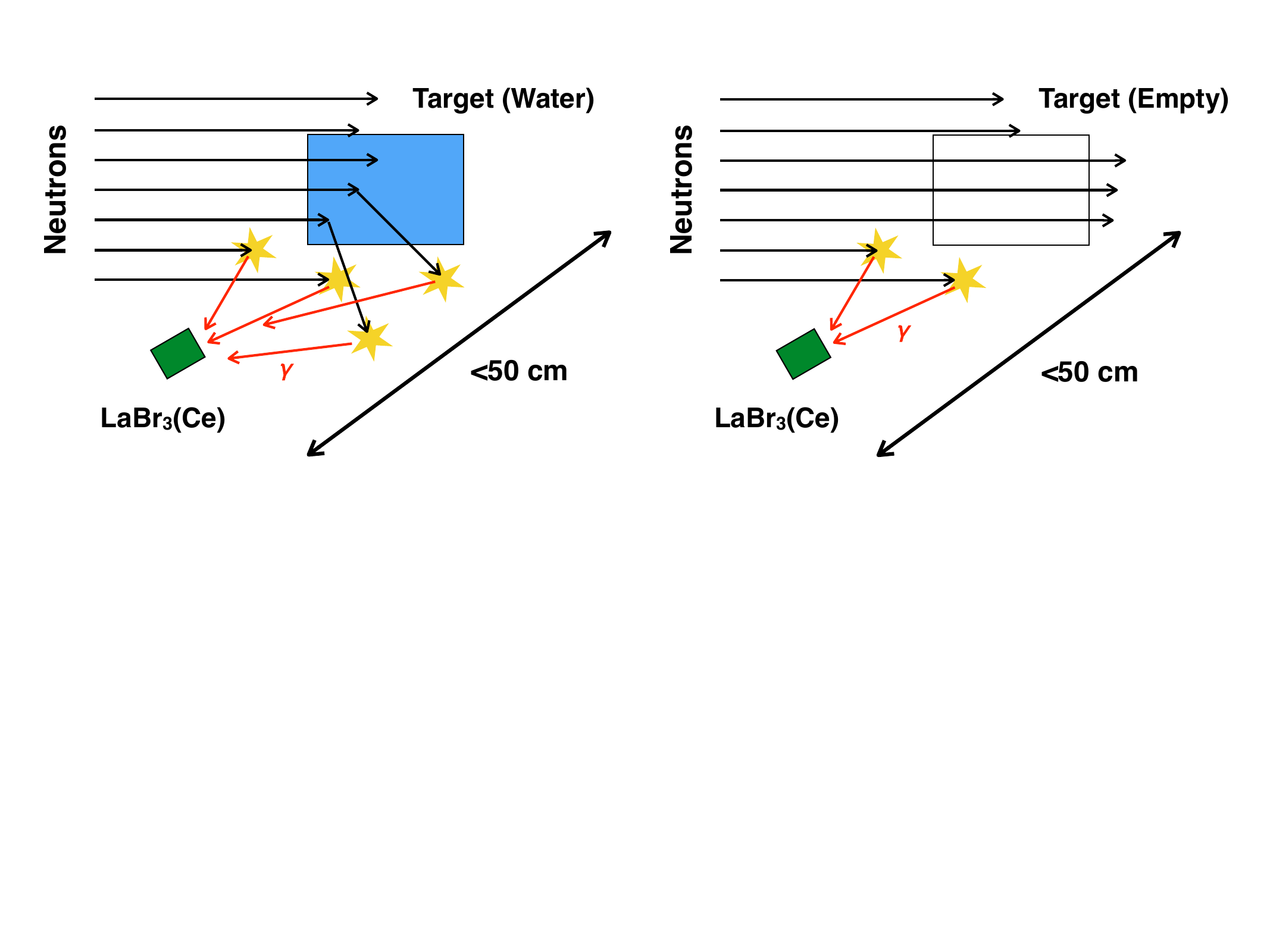}
  \end{center}
  \vspace{-180truept}
  \caption{Schematic illustration of the continuous background caused by
           $\gamma$-rays from scattered fast neutron reactions with the surrounding materials
           for a water-filled vessel (left) and an empty vessel (right).}
  \label{fig:schematic_fastnbkg}
  \end{figure*}

\begin{figure*}[htbp]
  \begin{center}
   \includegraphics[clip,width=18.5cm]{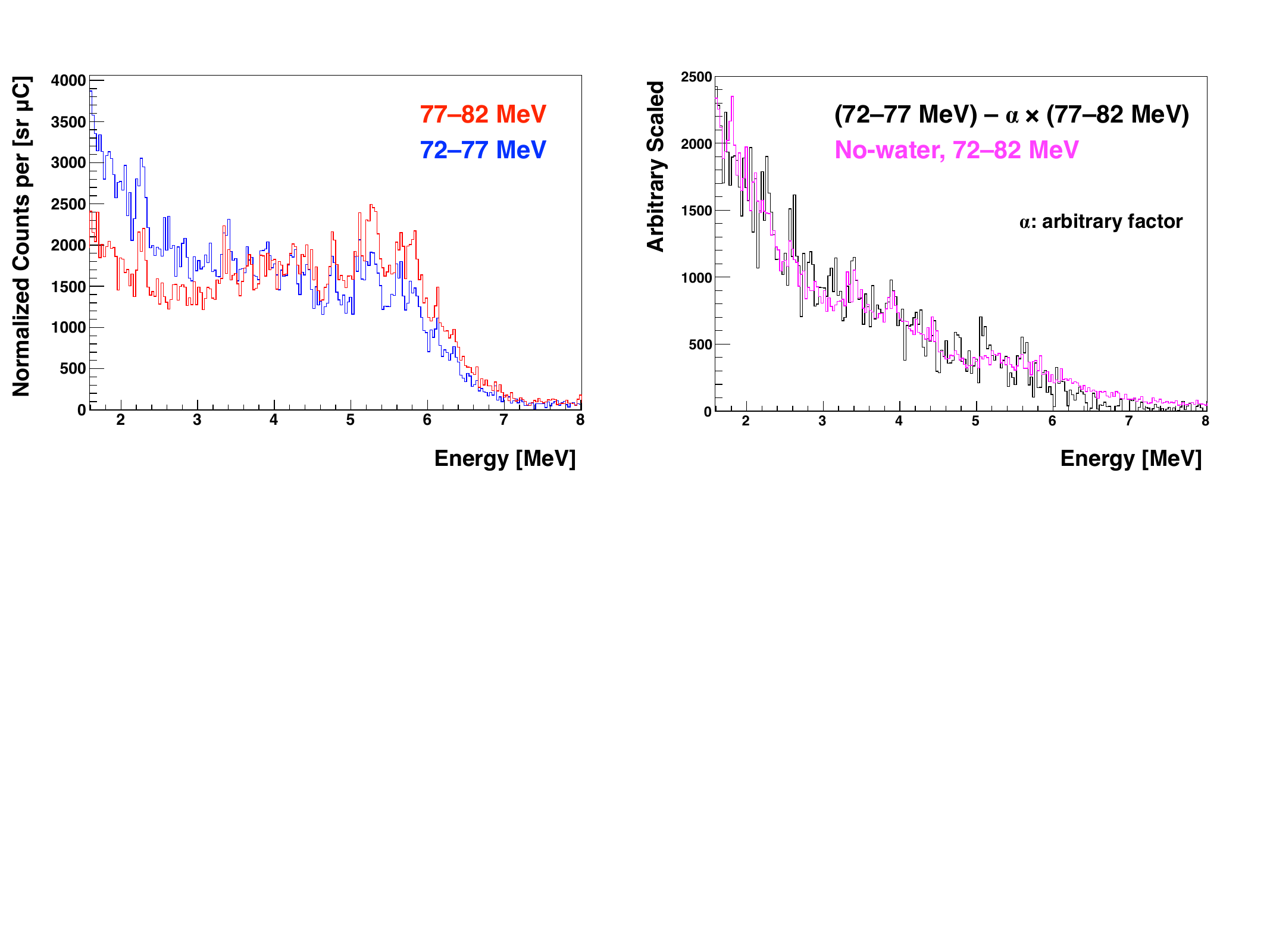}
  \end{center}
  \vspace{-210truept}
  \caption{Comparison of the spectra for the neutron energy regions 72$-$77~MeV and 77$-$82~MeV
           (left) and comparison between the shape difference of these two and the non-water
           spectrum for the 72$-$82~MeV region (right).}
  \label{fig:fastslowspectra}
  \end{figure*}

A continuous component remains in the spectrum after subtraction of the non-water background 
and the off-timing background. 
This is likely due to $\gamma$-rays which are produced from scattered neutron reactions 
with the surrounding materials, as depicted in the left panel of Figure~\ref{fig:schematic_fastnbkg}.
Those $\gamma$-rays are expected to come later than $\gamma$-rays which are emitted from 
the neutron-water reaction by the time between the neutron scattering in water and production 
of the $\gamma$-ray outside water. 
In the present experimental setup, the delay size is expected to be a few to 10~ns assuming 
the rescattering point is less than $\approx$100~cm from the water sample.
Hence the continuous component is predicted to be smaller in the spectrum with the TOF cut 
selecting faster neutrons. 
To confirm this, spectra from the different timing regions, corresponding to 72$-$77~MeV 
and 77$-$82~MeV in neutron kinetic energy, are compared. 
The 72$-$77~MeV region is $\approx$3~ns later than the 77$-$82~MeV region in TOF.
The result appears in the left panel of Figure~\ref{fig:fastslowspectra} and shows a 
clear difference in shape between these two spectra, as expected. 
The spectrum with an empty vessel and the on-timing TOF cut can be used as a template 
for the continuous background because the $\gamma$-ray source is expected to be the same,
as is schematically shown in the right panel of Figure~\ref{fig:schematic_fastnbkg}.
To check this the on-timing spectrum with an empty vessel is compared with 
the shape difference between the 72$-$77~MeV and 77$-$82~MeV spectra. 
Here the 77$-$82~MeV spectrum is subtracted from the 72$-$77~MeV spectrum with
an arbitrary scaling.
The result is given in the right panel of Figure~\ref{fig:fastslowspectra}, showing 
that the shapes of the two are similar.
Therefore the spectrum with an empty container and the on-timing cut can be used to predict 
the background spectrum from scattered fast neutrons with proper normalization. 
In the next section, this is used as a template spectrum in the fitting. 

\section{Cross Section Measurement}
\label{sec:xsection}

In Section~\ref{sec:bkgestimate}, the non-water and off-timing backgrounds are subtracted.
Now the observed spectrum is composed of signal (the $\gamma$-rays from neutron-oxygen reactions)
and the continuous background.
Using signal and background templates a fit to the observed data is performed to extract $\gamma$-ray
production cross sections.
The signal templates were made using a simulation based on the GEANT4 package \cite{geant4}, and
the continuous background template is obtained from the on-timing non-water data.

\subsection{Signal and continuous background templates}

In the GEANT4 simulation, the ${\rm LaBr_3(Ce)}$ detector and the acrylic container filled
with water are described and $\gamma$-rays are generated in the water. 
The generation point perpendicular to the beam axis follows the neutron beam profile as shown 
in Figure~\ref{fig:fluxprofile} and that in the direction parallel to the beam is determined by 
an exponential function based on the neutron mean free path in water. 
Here isotropic $\gamma$-ray emission is assumed. 
The simulated ${\rm LaBr_3(Ce)}$ spectrum is then smeared by the detector resolution. 
The resolution curve was obtained by fitting calibration points with 
$\sigma_{E}/E = p0 + p1/\sqrt{E} + p2/E$ ($E$ [MeV], $p0 \geq 0$) where $\sigma_E$ is 
the peak width determined by Gaussian fitting and $E$ is the peak energy in MeV.
The parameters are as follows: $p0 = (0.00+1.76) \times 10^{-4}$, 
$p1 = (1.29\pm0.04) \times 10^{-2}$, and $p2 = (-5.12\pm3.31) \times 10^{-4}$.
The Doppler effect should be considered for some peaks.
The size of this effect is expected to be $\approx$1\%~\cite{knoll} so 1\% is added to
the resolution obtained by the curve for the Doppler shifted peaks, 
7.12, 6.92, 6.32, 5.18, 4.44, 3.68, and 2.30/2.31~MeV. 
The consistency is cross checked using the 4.44~MeV peak in the ${\rm ^{241}Am/Be}$ calibration data.  
Note that the 2.30~MeV spectrum is used for the 2.30~MeV and 2.31~MeV peaks since
they cannot be differentiated with the ${\rm LaBr_3(Ce)}$.

As described in Section~\ref{sec:bkgestimate}, $\gamma$-rays from scattered neutron reactions
on the surrounding materials form the continuous background.
The shape of this background is obtained from the on-timing data with an empty vessel. 
In order to obtain a smooth shape template, the on-timing non-water spectrum is fit with 
an exponential function and the obtained function is used as a template.

\subsection{Spectrum fitting}

To fit the data with the templates above, a $\chi^2$ is calculated by comparing 
the observation and prediction ($=$ signal $+$ background):

  \begin{eqnarray}
   \chi^{2} = \sum_{i} \chi^{2}_{i}
            = \sum_{i} \left( \frac{N_{i}^{\rm obs} - N_{i}^{\rm pred}}{\sigma_{i}} \right)^{2}, \\ [+5pt]
   N_{i}^{\rm pred} = f_{0} \cdot N_{i}^{\rm bkg} + \sum_{j} f_{j} \cdot N_{i}^{{\rm sig}, j}.
  \label{eq:psdparam}
  \end{eqnarray}

\begin{figure*}[htbp]
   \begin{center}
    \includegraphics[clip,width=18.0cm]{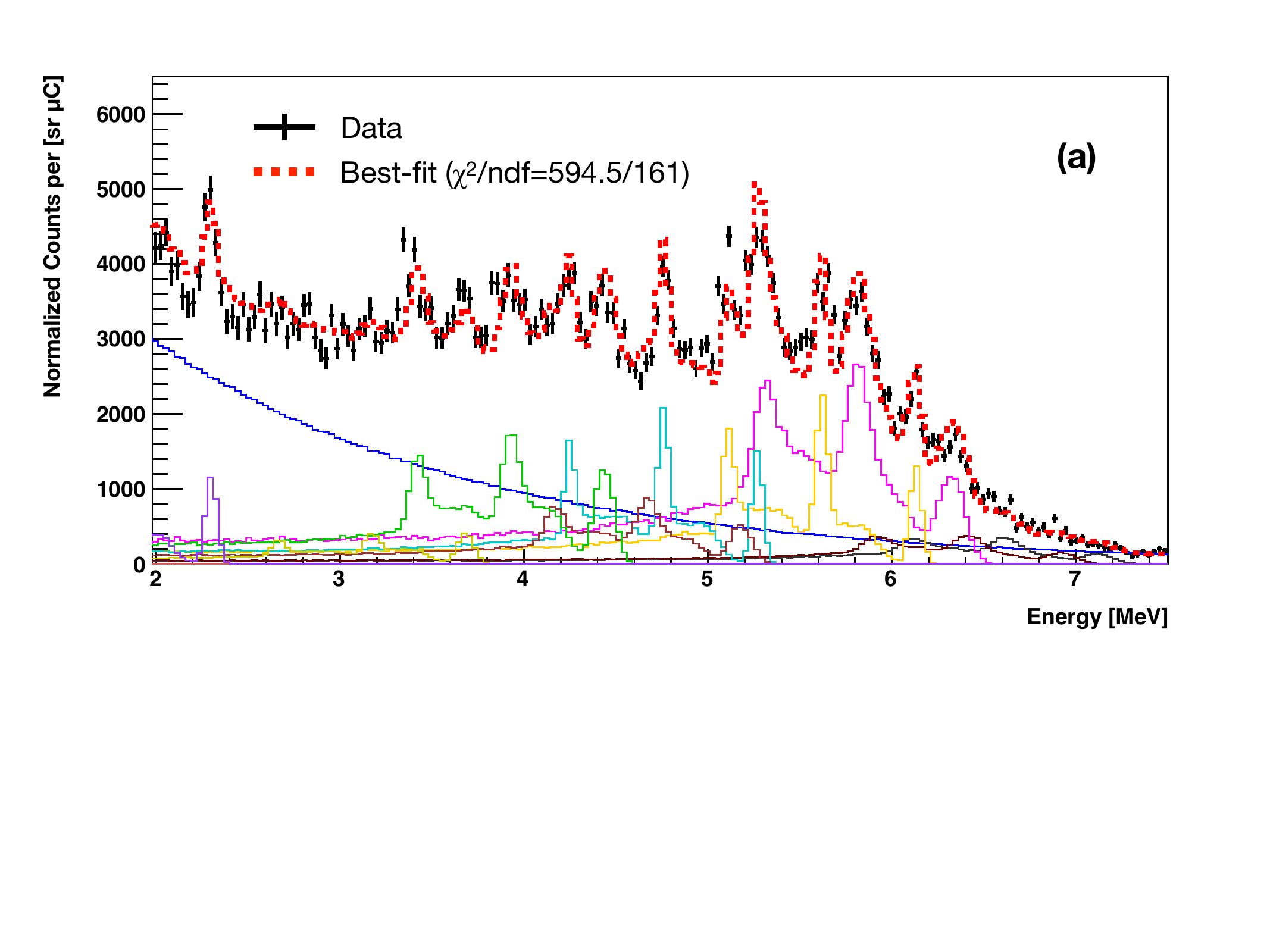}
   \end{center}
  \vspace{-150truept}
   \begin{center}
    \includegraphics[clip,width=18.0cm]{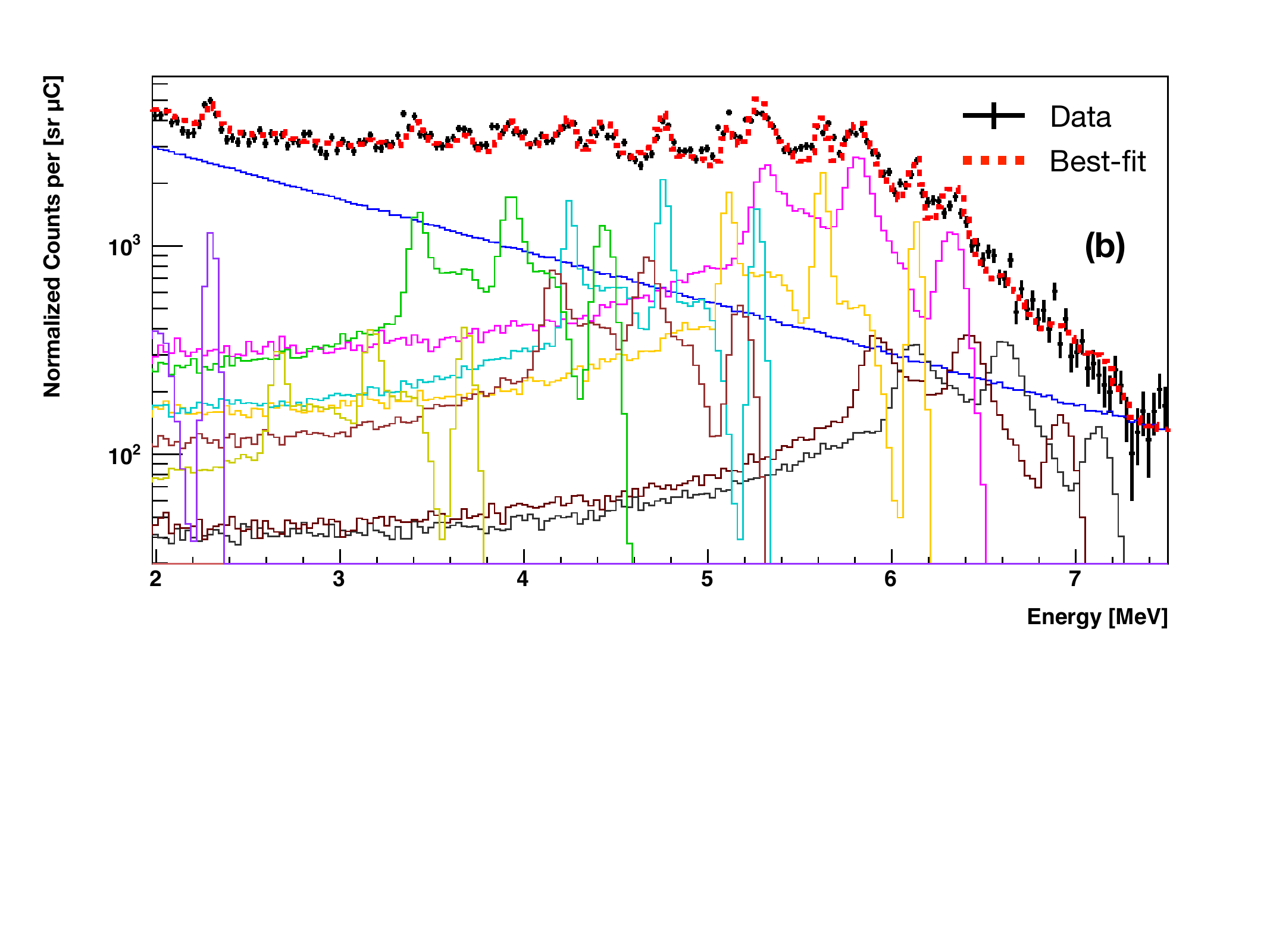}
   \end{center}
  \vspace{-120truept}
  \caption{Energy spectrum of the ${\rm LaBr_3(Ce)}$ scintillator after the TOF cut and
           after subtracting non-water and off-timing spectra (black points).
	       The best-fit spectrum (red dashed line), and the spectra of both the signal and background components
	       are shown in a linear (a) and a logarithmic (b) scale.
           Different colored spectra correspond to the fitted signals and continuous background:
           7.12~MeV (gray), 6.92~MeV (brown), 6.32~MeV (magenta), 6.13~MeV (orange),
           5.27~MeV (cyan), 5.18~MeV (dark red), 4.44~MeV (green), 3.68~MeV (yellow), 
	       2.30~MeV (violet), and continuous background (blue).}
  \label{fig:labrspectrum_fit}
  \end{figure*}

\noindent
Here $N_{i}^{\rm obs}$ [in units of ${\rm (sr \ \mu C)^{-1}}$] and $N_{i}^{\rm pred}$ represent the numbers of observed and predicted
events in the $i$-th energy bin, $N_{i}^{\rm pred}$ is the sum of the background and signal events
multiplied by the scale factors $f_j$ ($j=0, 1, \cdots, 9$; 0: background; 1: 7.12~MeV; 2: 6.92~MeV; 
3: 6.32~MeV; 4: 6.13~MeV; 5: 5.27~MeV; 6: 5.18~MeV; 7: 4.44~MeV; 8: 3.68~MeV; 9: 2.30~MeV).
The scale factors are normalized by the number of generated MC events ($10^8$ ${\rm (sr \ \mu C)^{-1}}$).
The error for the $i$-th energy bin ($\sigma_{i}$) considers the statistical uncertainties of
data and MC, the MC modeling error, and the energy resolution error.
The MC modeling was checked using $\gamma$-ray calibration sources and the absolute difference in
the number of detected events between data and MC is found to be 3.4\%. 
This check was performed at different distances and the obtained efficiency including detector acceptance looks reasonable with the extrapolated efficiency from the result for NaI(Tl) in Ref.~\cite{cecil}.
This difference is taken as an additional systematic error.
The energy resolution error is taken as the maximum difference in the number of events
in a bin between the nominal and either $\pm1\sigma$. 
The scale factors, $f_j$, are determined by minimizing the $\chi^2$ value.
Fitting is performed in two steps in order to isolate the high energy peaks and save computing time by 
reducing the number of parameters that need to be minimized simultaneously.
First, only the high energy region from 5.5 to 7.3~MeV is fit with only the background, 7.12~MeV,
6.92~MeV, 6.32~MeV, and 6.13~MeV signal spectra.
The second stage of the fit is performed for the energy range between 2.2 to 7.3~MeV 
with the scale factors for these four signals fixed while allowing the background to vary within 
the $\pm 2\sigma$ region found in the first step.
The fitting results are summarized in Table~\ref{tab:spectrumfit}.
The best-fit spectrum is shown together with the observed data in Figure~\ref{fig:labrspectrum_fit} 
and agrees well with the data.

  \begin{table}[htbp]
  \begin{center}
  \caption{Results of the first and second stages of spectrum fitting.}
  \label{tab:spectrumfit}
  %
   \begin{tabular}{l c | l c} \hline \hline
    First fitting & & Second fitting & \\ \hline
    $\chi^{2}$/ndf & $192.01/55 \approx 3.49$ & $\chi^{2}$/ndf & $594.49/161 \approx 3.69$ \\ \hline
    $f_0$ & $0.061 \pm 0.007$ & $f_0$ & $0.054 \pm 0.001$ \\
    $f_1$ & $0.14 \pm 0.01$   & $f_5$ & $0.40 \pm 0.02$   \\
    $f_2$ & $0.15 \pm 0.02$   & $f_6$ & $0.27 \pm 0.01$   \\
    $f_3$ & $0.94 \pm 0.03$   & $f_7$ & $0.49 \pm 0.02$   \\
    $f_4$ & $0.46 \pm 0.02$   & $f_8$ & $0.11 \pm 0.01$   \\ 
          &                   & $f_9$ & $0.14 \pm 0.02$   \\ \hline \hline
   \end{tabular}
  \end{center}
  \end{table}

\subsection{Cross sections for $\gamma$-ray production}

The $\gamma$-ray production cross section ($\sigma_{\gamma}^j$) for the $j$-th
signal is calculated as:

  \begin{eqnarray}
   \sigma_{\gamma}^j &=& \frac{N_{\rm fit}^{j}}{\epsilon_\gamma^{j} \cdot \phi_n \cdot T}
                     = f_{j} \times \frac{N_{\rm MC, generated}}{\phi_n \cdot T}, \\ [+5pt]
   N_{\rm fit}^{j} &=& f_{j} \cdot N_{\rm MC, detected}^{j}, \\ [+5pt]
   \epsilon_\gamma^{j} &=& \frac{N_{\rm MC, detected}^{j}}{N_{\rm MC, generated}}, 
  \label{eq:gammaxsec}
  \end{eqnarray}
  \vspace{1truept}

\noindent
where $\phi_n$ denotes the neutron flux [in units of ${\rm (sr \ \mu C)^{-1}}$], $T = 8.19 \times 10^{23} \ {\rm cm^{-2}}$ 
is the number of sample oxygen nuclei per unit area, $\epsilon_\gamma$ is the $\gamma$-ray detection 
efficiency including detector acceptance (e.g., the efficiency including detector acceptance estimated by MC for the 6.13~MeV, 5.27~MeV, and 4.44~MeV $\gamma$-rays are 0.0049\%, 0.0061\%, and 0.0053\%, respectively), and $N_{\rm MC, generated}$ ($N_{\rm MC, detected}$) is the number of generated (detected) events in the GEANT4 simulation.
In the simulation $10^8$ events are generated for every peak. 
There are contributions from low energy neutrons produced in the scattering of initial neutrons inside water. 
The effect of this multiple scattering is expected to be sizable since the $\gamma$-ray production cross section is 
usually higher at lower energies, even though the flux of scattered neutrons is smaller than the initial neutron flux 
in Figure~\ref{fig:fluxdist}, therefore cross sections need to be corrected. 
This effect was evaluated using an MC simulation, where 80-MeV neutrons are injected in a water volume of 
the same size as our sample.
Here the neutron flux after the scattering was multiplied with the cross section ratio between each energy to 80~MeV,
which was taken from Ref.~\cite{nelson}, and the relative target mass is considered as the ratio between 
distance to the sample surface after the scattering to height of the water sample. 
Then the convolution was performed along neutron energies to obtain the relative effect. 
The results show that the effect is 35\%.
The contribution from tertiary or more scattered neutrons is small. 
A factor of 0.65 is then taken to correct $\gamma$-ray cross sections.

The systematic uncertainty on the cross section is composed of errors from the spectrum fitting, 
the neutron flux estimation, the lower energy neutron contribution due to multiple scattering, and 
the estimation of non-water background from the vessel's back-face, as summarized in Table~\ref{tab:gammaxsecsyst}. 
The first two sources were detailed in the former parts of the article and the rest is explained in the following part. 

The systematic uncertainty about the correction factor for low energy neutron contributions has two sources, 
the uncertainty of neutron reaction model used in the MC simulation and differential cross section shape in Ref.~\cite{nelson}.
The former was evaluated by changing neutron cross sections by $\pm$30\% as an assigned error on this model. 
This gives a $+$5\%/$-$38\% change in the correction factor. 
The latter uncertainty was evaluated by changing the functional form to accommodate relative difference 
in differential cross section shape between $\gamma$-ray peaks in Ref.~\cite{nelson}. 
This produces a $\approx$31\% change in the correction factor. 
In the end, the systematic uncertainty for this effect is $+$31\%/$-$49\% (a correction factor is $0.65^{+0.20}_{-0.32}$).

Another uncertainty is about the non-water background estimation. 
In the water-filled measurement $\approx$56\% of neutrons do not reach the back-face
of the acrylic vessel based on the neutron mean free path in water and the vessel's length.
This may lead to an overestimate of the background from the non-water measurement 
since there is no such neutron deficit at the back-face of the acrylic vessel in that case. 
Since the ratio of the volume of the acrylic vessel's back-face to its total is $\approx$23\% and 
the neutron flux at the radial position of the acrylic barrel is $\approx$1.7 times smaller than at 
the beam center (c.f. Figure~\ref{fig:fluxprofile}), the effective contribution of neutron-induced 
events on back-face of the vessel is about $\approx$29\% of the non-water rate.
Based on Figure~\ref{fig:labrspectrum_tof} the contribution of neutrons from the acrylic vessel 
(the non-water line) to the spectrum with water is at most 50\% above 6.8~MeV and 30\% below. 
Therefore, the maximum impact of the reduced flux at the back-face of the acrylic is given by 
the product of these factors: 8\% for the 7.12~MeV and 6.92~MeV peaks and 5\% for the others. 
These quantities are taken as systematic uncertainties. 
 
The measured cross section for each $\gamma$-ray is summarized in Table~\ref{tab:gammaxsecresult}.
The result for the 2.30~MeV and 2.31~MeV peaks is inclusive, since the two cannot be separated 
in the current measurement.
Here the uncertainties are calculated by adding all the sources explained above in quadrature. 
Note that the asymmetric uncertainties stem from the multiple scattering error. 

  %

  \begin{table}[htbp]
  \begin{center}
  \caption{Systematic uncertainties in the $\gamma$-ray cross section measurement.
           Statistical uncertainties in the observed data are included in the fitting errors and 
	   their impacts are only subdominant.
	   Note that ``Fitting'' is for the error from the spectrum fitting, ``Flux'' for the neutron flux estimation, 
	   ``Multiple scattering'' for the lower energy neutron effect, and ``Back-face'' for the non-water background estimation.}
  \label{tab:gammaxsecsyst}
  \vspace{2truept}
   \begin{tabular}{c c c c c} \hline \hline
    $E_{\gamma}$ [MeV] & Fitting & Flux & Multiple scattering & Back-face \\ \hline
    7.12      & $\pm7$\%  & $\pm11$\% & $+$31\%/$-$49\% & $\pm8$\% \\ 
    6.92      & $\pm13$\% & $\pm11$\% & $+$31\%/$-$49\% & $\pm8$\% \\ 
    6.32      & $\pm3$\%  & $\pm11$\% & $+$31\%/$-$49\% & $\pm5$\% \\ 
    6.13      & $\pm4$\%  & $\pm11$\% & $+$31\%/$-$49\% & $\pm5$\% \\ 
    5.27      & $\pm5$\%  & $\pm11$\% & $+$31\%/$-$49\% & $\pm5$\% \\ 
    5.18      & $\pm4$\%  & $\pm11$\% & $+$31\%/$-$49\% & $\pm5$\% \\ 
    4.44      & $\pm4$\%  & $\pm11$\% & $+$31\%/$-$49\% & $\pm5$\% \\ 
    3.68      & $\pm9$\%  & $\pm11$\% & $+$31\%/$-$49\% & $\pm5$\% \\ 
    2.30/2.31 & $\pm14$\% & $\pm11$\% & $+$31\%/$-$49\% & $\pm5$\% \\ \hline \hline
   \end{tabular}
  \end{center}
  \end{table}


  \begin{table}[htbp]
  \begin{center}
  \caption{Results of the $\gamma$-ray production cross sections.}
  \label{tab:gammaxsecresult}
  \vspace{2truept}
   \begin{tabular}{c c} \hline \hline
    $E_{\gamma}$ [MeV] & $\sigma_{\gamma}$ [mb] \\ \hline
    7.12      & $0.6^{+0.2}_{-0.3}$ \\
    6.92      & $0.6^{+0.2}_{-0.3}$ \\
    6.32      & $3.8^{+1.3}_{-1.9}$ \\
    6.13      & $1.9^{+0.6}_{-1.0}$ \\
    5.27      & $1.6^{+0.5}_{-0.8}$ \\
    5.18      & $1.1^{+0.4}_{-0.6}$ \\
    4.44      & $2.0^{+0.7}_{-1.0}$ \\
    3.68      & $0.5^{+0.2}_{-0.3}$ \\
    2.30/2.31 & $0.6^{+0.2}_{-0.3}$ \\ \hline \hline
   \end{tabular}
  \end{center}
  \end{table}

\section{Discussion}
\label{sec:discussion}

\begin{figure}[htbp]
  \begin{center}
   \includegraphics[clip,width=9.5cm]{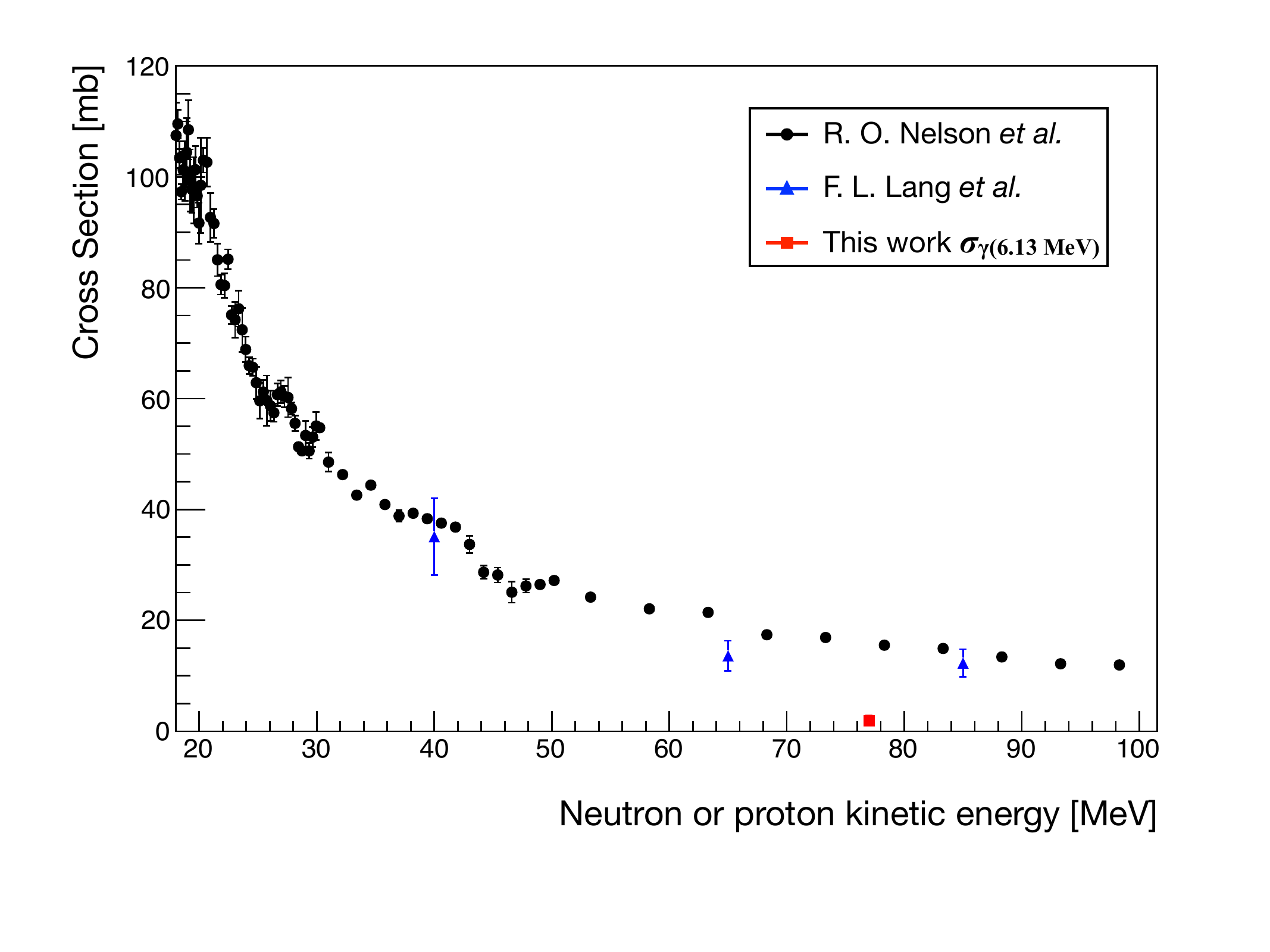}
  \end{center}
  \vspace{-30truept}
  \caption{Comparison of the measured 6.13~MeV $\gamma$-ray production cross sections. 
           This work is shown by a red square while
	   the results of Ref.~\cite{nelson} appear as black circles 
	   and those of Ref.~\cite{lang} are shown by blue triangles.
	   Note that the latter is based on measurements of ${\rm ^{16}O}(p,p'\gamma)$.}
  \label{fig:comparison613}
  \end{figure}

\begin{figure}[htbp]
  \begin{center}
   \includegraphics[clip,width=9.5cm]{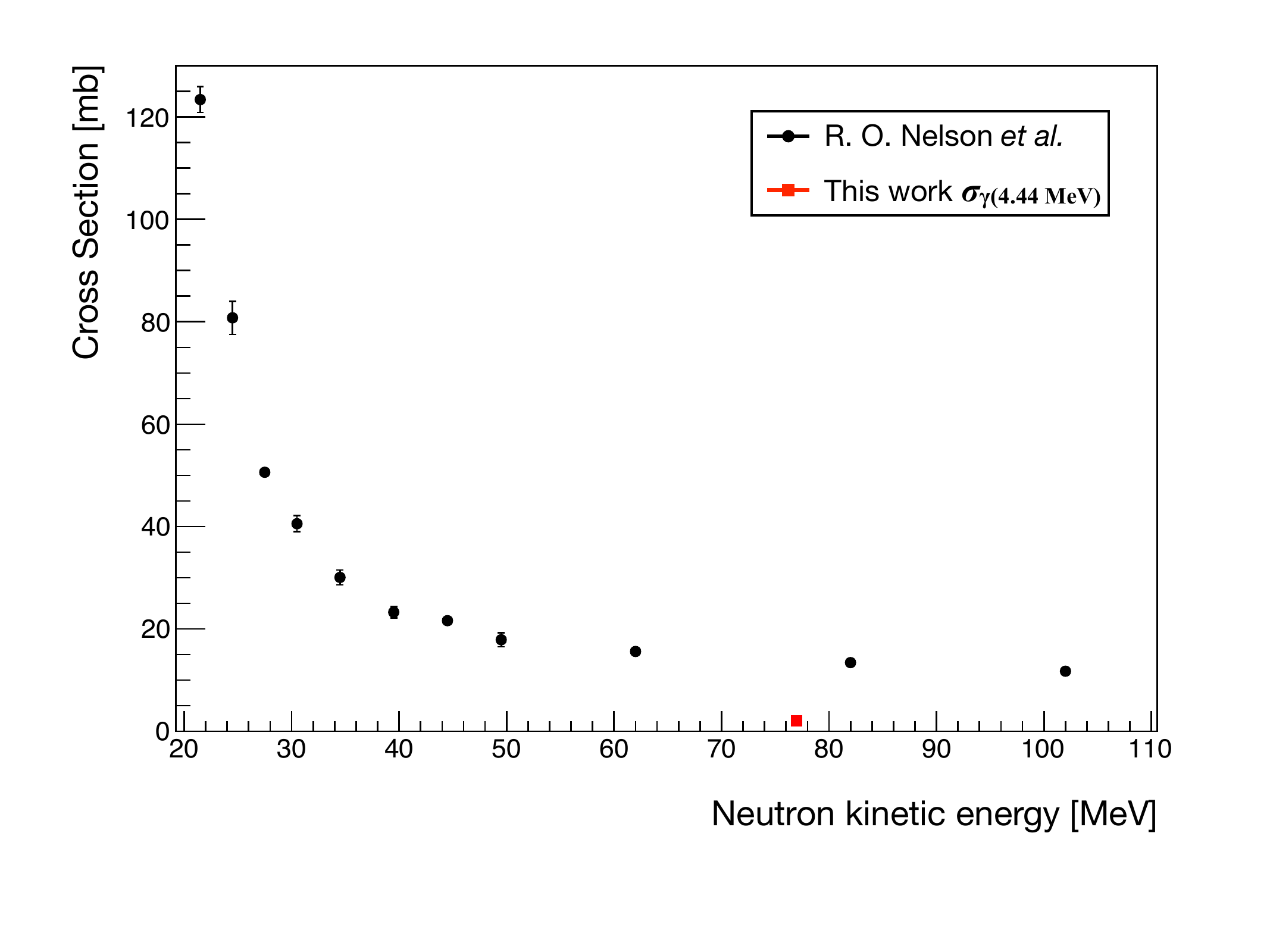}
  \end{center}
  \vspace{-30truept}
  \caption{Comparison of the measured 4.44~MeV $\gamma$-ray production cross sections on oxygen. 
           The present result is shown by a red square while the results of 
	   Ref.~\cite{nelson} appear as black circles.} 
  \label{fig:comparison444}
  \end{figure}

The measured cross sections are the first results using a monoenergetic neutron beam of 77~MeV.
The result for the 6.32~MeV $\gamma$-ray is the first measurement in this energy region. 
Figure~\ref{fig:comparison613} compares the current result for the 6.13~MeV $\gamma$-ray 
cross section with similar measurements~\cite{nelson,lang}.
Unlike this result, Ref.~\cite{nelson} used a broadband neutron beam and was based on a counting method.
The result from Ref.~\cite{lang} is for the ${\rm ^{16}O}(p,p')$ reaction.
%
Measurements of the 5.27~MeV and 4.44~MeV $\gamma$-ray cross sections are 
also presented in Ref.~\cite{nelson}, both of which are larger than the results of this work 
at similar neutron energies.
Figure~\ref{fig:comparison444} shows a comparison of the 4.44~MeV $\gamma$-ray production cross sections.
Further details can be obtained in the nuclear library EXFOR~\cite{exfor}. 
Unfortunately, in spite of a lot of careful checks, it is not fully understood why the results from this work show smaller cross sections than the ones previously reported at around similar energies. Further careful investigation is needed by providing more data points as well as theoretical calculations.

The results presented here provide valuable inputs to the modeling of $\gamma$-ray 
production via neutron-${\rm ^{16}O}$ reactions. 
Such reactions are of particular relevance to neutrino-${\rm ^{16}O}$ neutral-current 
scattering measurements in water Cherenkov detectors~\cite{t2kncqe1to3,t2kncqe1to9,skncqe}.
Similarly, these data are expected to be beneficial to water Cherenkov experiments 
seeking to measure the final state neutron multiplicity of neutrino interactions,
such as the ANNIE experiment~\cite{annie1,annie2}, because understanding neutron transport 
and subsequent $\gamma$-ray production are essential for identifying the signal.

\section{Conclusion}
\label{sec:conclusion}

A measurement of $\gamma$-ray emission from neutron-${\rm ^{16}O}$ reactions was 
carried out at RCNP using a nearly monoenergetic neutron beam with a mean energy of 77~MeV.
In the experiment $\gamma$-rays were measured using a ${\rm LaBr_3(Ce)}$ scintillator
and other dedicated measurements, to understand the incident neutron flux and expected 
backgrounds, were performed using other detectors.
The production cross sections were measured for nine $\gamma$-ray components varying from 2.3 to 7.12~MeV. 
These are the first measurement results at 77~MeV using a monoenergetic beam.
The measurements presented here will be of use for developing neutron interaction models, 
which are particularly important for understanding neutron-induced $\gamma$-ray productions
at water Cherenkov detectors.

\section*{Acknowledgment}

The authors are grateful to the RCNP staff for giving a beam time to the experiment 
and the accelerator group for supplying a stable beam. 
We also thank Dr. Y. Iwamoto (JAEA) and Dr. S. Meigo (JAEA) 
for their helpful advices on neutron flux measurements. 
This work was supported by Japan MEXT KAKENHI Grants No. 17J06141, No.26400292, 
No. 25105002, No. 18H05537, and No. JP20H00162.



\begin{thebibliography}{0}%
\makeatletter
\providecommand \@ifxundefined [1]{%
 \@ifx{#1\undefined}
}%
\providecommand \@ifnum [1]{%
 \ifnum #1\expandafter \@firstoftwo
 \else \expandafter \@secondoftwo
 \fi
}%
\providecommand \@ifx [1]{%
 \ifx #1\expandafter \@firstoftwo
 \else \expandafter \@secondoftwo
 \fi
}%
\providecommand \natexlab [1]{#1}%
\providecommand \enquote  [1]{``#1''}%
\providecommand \bibnamefont  [1]{#1}%
\providecommand \bibfnamefont [1]{#1}%
\providecommand \citenamefont [1]{#1}%
\providecommand \href@noop [0]{\@secondoftwo}%
\providecommand \href [0]{\begingroup \@sanitize@url \@href}%
\providecommand \@href[1]{\@@startlink{#1}\@@href}%
\providecommand \@@href[1]{\endgroup#1\@@endlink}%
\providecommand \@sanitize@url [0]{\catcode `\\12\catcode `\$12\catcode
  `\&12\catcode `\#12\catcode `\^12\catcode `\_12\catcode `\%12\relax}%
\providecommand \@@startlink[1]{}%
\providecommand \@@endlink[0]{}%
\providecommand \url  [0]{\begingroup\@sanitize@url \@url }%
\providecommand \@url [1]{\endgroup\@href {#1}{\urlprefix }}%
\providecommand \urlprefix  [0]{URL }%
\providecommand \Eprint [0]{\href }%
\providecommand \doibase [0]{http://dx.doi.org/}%
\providecommand \selectlanguage [0]{\@gobble}%
\providecommand \bibinfo  [0]{\@secondoftwo}%
\providecommand \bibfield  [0]{\@secondoftwo}%
\providecommand \translation [1]{[#1]}%
\providecommand \BibitemOpen [0]{}%
\providecommand \bibitemStop [0]{}%
\providecommand \bibitemNoStop [0]{.\EOS\space}%
\providecommand \EOS [0]{\spacefactor3000\relax}%
\providecommand \BibitemShut  [1]{\csname bibitem#1\endcsname}%
\let\auto@bib@innerbib\@empty
\end{thebibliography}%


\begin{thebibliography}{100}

\bibitem{superk}
  S. Fukuda $et \ al.$ (Super-Kamiokande Collaboration),
  Nucl. Instr. and Meth. Phys. Res. A 501 (2003).
\bibitem{skgd}
  J. Beacom and M. Vagins,
  Phys. Rev. Lett. 93, 171101 (2004).
\bibitem{hyperk}
  K. Abe $et \ al.$ (Hyper-Kamiokande Proto-Collaboration),
  arXiv:1805.04163 (2018).
\bibitem{sksrn12}
  M. Malek $et \ al.$ (Super-Kamiokande Collaboration),
  Phys. Rev. Lett. 90, 061101 (2003).
\bibitem{sksrn123}
  K. Bays $et \ al.$ (Super-Kamiokande Collaboration),
  Phys. Rev. D 85, 052007 (2012).
\bibitem{sksrn4}
  H. Zhang $et \ al.$ (Super-Kamiokande Collaboration),
  Astropart. Phys. 60, 41 (2015).
\bibitem{sksrn1234}
  K. Abe $et \ al.$ (Super-Kamiokande Collaboration),
  Phys. Rev. D 104, 122002 (2021).
\bibitem{t2kdm1}
  P. deNiverville $et \ al.$,
  Phys. Rev. D 86, 035022 (2012).
\bibitem{t2kdm2}
  P. deNiverville $et \ al.$,
  Phys. Rev. D 95, 035006 (2017).
\bibitem{t2ksterile}
  K. Abe $et \ al.$ (T2K Collaboration),
  Phys. Rev. D 99, 071103(R) (2019).
\bibitem{ankowski}
  Arthur M. Ankowski $et \ al.$,
  Phys, Rev. Lett. 108, 052505 (2012).
\bibitem{t2kexp}
  K. Abe $et \ al.$ (T2K Collaboration),
  Nucl. Instr. and Meth. Phys. Res. A, 659 (2011).
\bibitem{t2kncqe1to3}
  K. Abe $et \ al.$ (T2K Collaboration),
  Phys. Rev. D 90, 072012 (2014).
\bibitem{t2kncqe1to9}
  K. Abe $et \ al.$ (T2K Collaboration),
  Phys. Rev. D 100, 112009 (2019).
\bibitem{gcalor1}
  C. Zeitnitz and T. A. Gabriel,
  Nucl. Instr. and Meth. Phys. Res. A 349 (1994).
\bibitem{gcalor2}
  C. Zeitnitz and T. A. Gabriel,
  Proceedings of the International Conference on Monte Carlo
  Simulation in High Energy and Nuclear Physics (MC93) (1993).
\bibitem{endf}
  National Nucler Data Center,
  \url{https://www.nndc.bnl.gov/exfor/endf00.jsp}
\bibitem{inc1}
  J. Cugnon, C. Volant, and S. Vuillier, 
  Nucl. Phys. A 620, 4 (1997). 
\bibitem{inc2}
  A. Boudard, J. Cugnon, S. Leray, and C. Volant, 
  Phys. Rev. C 66, 044615 (2002).
\bibitem{rcnp1}
  T. Miura $et \ al.$,
  Proceedings of 13th International Conference on Cyclotrons
  and their Applications,
  Vancouver, Canada (1992).
\bibitem{rcnp2}
  T. Saito $et \ al.$,
  Proceedings of 14th International Conference on Cyclotrons
  and their Applications,
  Cape Town, South Africa (1995).
\bibitem{rcnp3}
  S. Ninomiya $et \ al.$,
  Proceedings of 17th International Conference on Cyclotrons
  and their Applications,
  Tokyo, Japan (2004).
\bibitem{scinfulqmd1}
  D. Satoh $et \ al.$,
  JAEA-DATA/CODE 2006-023 (2006).
\bibitem{scinfulqmd2}
  T. Kajimoto $et \ al.$,
  Nucl. Instr. and Meth. Phys. Res. A 665 (2011).
\bibitem{flux2}
  Y. Iwamoto $et \ al.$,
  Nucl. Instr. and Meth. Phys. Res. A 804 (2015).
\bibitem{scinful}
  J. K. Dickens,
  Technical Report ORTN-6436 (1988).
\bibitem{satoh}
  D. Satoh $et \ al.$,
  Journal of Nuclear Science and Technology, 43, 7 (2006).
\bibitem{nakao}
  N. Nakao $et \ al.$,
  Nucl. Instr. and Meth. Phys. Res. A 362 (1995).
\bibitem{flux1}
  S. Meigo $et \ al.$,
  Nucl. Instr. and Meth. Phys. Res. A 401 (1997).
\bibitem{exfor}
  Experimental Nuclear Reaction Data (EXFOR),
  \url{https://www-nds.iaea.org/exfor/137500}
\bibitem{nelson}
  R. O. Nelson $et \ al.$,
  Journal of Nuclear Science and Engineering, 138 (2001).
\bibitem{ejiri}
  H. Ejiri,
  Phys. Rev. C 48, 1442 (1993).
\bibitem{leuschner}
  M. Leuschner $et \ al.$,
  Phys. Rev. C 49, 955 (1994).
\bibitem{dimbylow}
  P. J. Dimbylow,
  Phys. Med. Biol., 25, 637 (1980).
\bibitem{csipsd}
  Y. Ashida $et \ al.$,
  Prog. Theor. Exp. Phys. 043H01 (2018).
  Erratum: Prog. Theor. Exp. Phys. 2018, 069201 (2018).
\bibitem{geant4}
  S. Agostinelli $et \ al.$,
  Nucl. Instr. and Meth. Phys. Res. A 506 (2003).
\bibitem{knoll}
  G. F. Knoll,
  John Wiley \& Sons (2001).
\bibitem{cecil}
  F. E. Cecil $et \ al.$,
  Nucl. Instr. and Meth. Phys. Res. A 234 (1985).
\bibitem{lang}
  F. L. Lang $et \ al.$,
  Phys. Rev. C 35, 4 (1987).
\bibitem{skncqe}
  L. Wan $et \ al.$ (Super-Kamiokande Collaboration),
  Phys. Rev. D 99, 032005 (2019).
\bibitem{annie1}
  I. Anghel $et \ al.$,
  arXiv:1504.01480 (2015).
\bibitem{annie2}
  A. R. Back $et \ al.$,
  arXiv:1707.08222 (2017).

\end{thebibliography}
\end{document}